\documentclass{elsart}

\usepackage{amsmath}
\usepackage{amssymb}
\usepackage[dvips]{graphicx}
\usepackage{color}

\newcommand{\phme}[2]{\langle
                      (1 \bar{3})_{#2}  |  #1  |
                      (4 \bar{2})_{#2}
                      \rangle}
\newcommand{\sgn}[1]{(-)^{#1}}
\newcommand{\cg}[6]{\langle #1 \, #2 \: #3 \, #4\ | \ #5 \, #6 \rangle}
\newcommand{\mmme}[1]{\langle
                      1 2  |  #1  |
                      3 4
                      \rangle}
\newcommand{\ppme}[2]{\langle
                      (1 2)_{#2}  |  #1  |
                      (3 4)_{#2}
                      \rangle}
\newcommand{\wsj}[6]{\left\{ \begin{array}{ccc}
                             #1 & #2 & #3 \\
                             #4 & #5 & #6
                             \end{array} \right\} }
\newcommand{\mc}[9]{\langle #1 \, #2 \: #3 \, #4 \: #9 \ | \
                            #5 \, #6 \: #7 \, #8 \: #9 \rangle}
\newcommand{\wnj}[9]{\left\{ \begin{array}{ccc}
                             #1 & #2 & #3 \\
                             #4 & #5 & #6 \\
                             #7 & #8 & #9
                             \end{array} \right\} }

\newcommand{\sprme}[3]{\langle #1  \; \| \;  #2  \; \| \; #3 \rangle}
\newcommand{\del}[2]{\delta_{#1 \, #2}}

\newcommand{\ck}[2]{C^{(#1)} (\hat{#2})}
\newcommand{\intr}[1]{\! \int_0 ^{\infty} \! [{#1}^2 \, {\rm d}#1]}

\newcommand{\ckq}[3]{C^{(#1)}_{#2} (\hat{#3})}
\newcommand{\Ykq}[3]{Y_{#1 #2} (\hat{#3})}

\newcommand{\intt}[1]{\! \int \! {\rm d}^3 #1}
\newcommand{\spjrme}[3]{\langle
                      (l_{#1} {\textstyle\frac{1}{2}})_{j_{#1}}
                      \; \| \;  #2  \; \| \;
                      (l_{#3} {\textstyle\frac{1}{2}})_{j_{#3}}
                      \rangle}
\newcommand{\Ffac}[3]{F(l_{#1} j_{#1}; l_{#2} j_{#2} ; #3)}
\newcommand{\lropme}[3]{\langle
                      l_{#1}
                      \; \| \;  #2  \; \| \;
                      l_{#3}
                      \rangle}
\newcommand{\Gfac}[4]{G(l_{#1} j_{#1}; l_{#2} j_{#2} ; #3 #4)}

\newcommand{\jjmeph}[1]{\langle
   (l_1 {\textstyle\frac{1}{2}})_{j_1} (l_3 {\textstyle\frac{1}{2}})_{\bar j_3} \: ; \:
   \lambda \mu
                      \; | \;  #1  \; | \;
   (l_4 {\textstyle\frac{1}{2}})_{j_4} (l_2 {\textstyle\frac{1}{2}})_{\bar j_2} \: ; \:
   \lambda \mu
                      \rangle}
\newcommand{\spjme}[3]{\langle
                      (l_{#1} {\textstyle\frac{1}{2}})_{j_{#1}}
                      \; \| \;  #2  \; \| \;
                      (l_{#3} {\textstyle\frac{1}{2}})_{j_{#3}}
                      \rangle}
\newcommand{\ckrme}[3]{\langle
                      l_{#1}
                      \; \| \;  C^{(#2)}  \; \| \;
                      l_{#3}
                      \rangle}
\newcommand{\grd}{\bigtriangledown}
\newcommand{\cka}[1]{C^{(#1)} (\hat r_1)}
\newcommand{\ckb}[1]{C^{(#1)} (\hat r_2)}
\newcommand{\cga}[6]{\langle
                      l_1
                      \; \| \;  u^{(#1 #2 , #3 #4)} (r_1,r_2) \
                      \left[ C^{(#5)} (\hat r_1)
                      \otimes \grd_{(1)} \right]^{(#6)} \; \| \;
                      l_3
                     \rangle}
\newcommand{\cgb}[6]{\langle
                      l_2
                      \; \| \;  u^{(#1 #2 , #3 #4)} (r_1,r_2) \
                      \left[ C^{(#5)} (\hat r_2)
                      \otimes \grd_{(2)} \right]^{(#6)} \; \| \;
                      l_4
                     \rangle}
\newcommand{\comm}[2]{[ \, #1 , #2 ] \, }
\newcommand{\cgga}[7]{\langle
                       l_1
                       \; \| \;  u^{(#1 #2 , #3 #4)} (r_1,r_2)
                       \left[ C^{(#5)} (\hat r_1)
                       \otimes
                       [ \grd_{(1)} \otimes \grd_{(1)} ]^{(#6)}
                       \right ]^{(#7)} \; \| \;
                       l_3
                      \rangle}
\newcommand{\cggb}[7]{\langle
                       l_2
                       \; \| \;  u^{(#1 #2 , #3 #4)} (r_1,r_2)
                       \left[ C^{(#5)} (\hat r_2)
                       \otimes
                       [ \grd_{(2)} \otimes \grd_{(2)} ]^{(#6)}
                       \right ]^{(#7)} \; \| \;
                       l_4
                      \rangle}

\begin{document}

\begin{frontmatter}

\title{Matrix elements of the Argonne $v_{18}$ potential}

\author{Bogdan Mihaila}
\ead{bmihaila@lanl.gov}
\address{
   Los Alamos National Laboratory,
   Los Alamos, NM 87545}

\begin{abstract}
   We discuss two approaches to the calculation of matrix elements of the
   Argonne $v_{18}$ potential. The first approach is applicable in the case of a single-particle basis of
   harmonic-oscillator wave functions.
   In this case we use the Talmi transformation, implemented numerically using the
   Moshinsky transformation brackets, to
   separate the center-of-mass and relative coordinates degrees of freedom.
   Integrals involving the radial part of the potential are performed
   using Gauss-Hermite quadrature formulas, and convergence is achieved
   for sets of at least 512~points.
   We validate the calculation of matrix elements of
   the Argonne $v_{18}$ potential
   using a second approach suitable for the case of an
   arbitrary functional form of the single-particle wave functions.
When the model space is represented in terms of harmonic-oscillator
wave functions, results obtained using these two approaches are shown to
to be identical within numerical accuracy.
\end{abstract}


\end{frontmatter}

\section{Introduction}

Direct comparison of experimental data and theory become ambiguous
if problems such as the many-body problem are not solved accurately.
This is particularly true in nuclear physics, where one uses
numerical solutions of the Schr\"odinger equation to constrain in
part the effective interaction between nucleons~\cite{urbana}.
Despite tremendous accomplishments in the past decade in carrying
out ab~initio studies of nuclear structure of light
systems~\cite{gfmc1,gfmc2,gfmc3,gfmc4,ncsm1,ncsm2,ncsm3,ncsm4,exps-1,exps-2,exps-3,expZ-1,expZ-2,expZ-3},
numerical solutions of medium and large nuclear systems are yet to
be performed in the framework of realistic nuclear
interactions~\cite{av18,cd-bonn,idaho}.

The matrix elements of the interaction represent the building blocks
of any many-body approach which shares features in common with the
shell model approach. Here, the model space is represented in terms
of a complete set of single-particle wave functions. All possible
single-particle configurations with the appropriate properties
should in principle be used to diagonalize the Hamiltonian, and
obtain the associated spectrum of eigenvalues and eigenfunctions.
However, the single-particle model space is in principle infinite.
Hence, finite reductions of the model space must be considered and
convergence to the continuum limit must be achieved. The efficiency
of this procedure is linked to the ability of the functional form of
the single-particle wave functions to capture the relevant
correlations at both short- and long-range scale.

Traditionally, matrix elements of the interaction are formed in a
basis of harmonic-oscillator wave functions. This has the advantage
that one can perform an exact separation of the center-of-mass
and relative coordinates  degrees of freedom using the Talmi transformation approach~\cite{Talmi}
(via its numerical implementation based on the Moshinsky transformation brackets~\cite{brackets}), and integrals can be accurately performed using
Gauss-Hermite quadrature formulas.
Therefore, we will by discussing the calculation of the two-body
matrix elements in a harmonic-oscillator basis. This is done both
for completeness, and for testing purposes, as the more elaborated
procedures to be discussed in the second part of this paper will require
validation.

Unfortunately, the asymptotic form of
the harmonic-oscillator wave functions is Gaussian which leads to
densities decaying as $e^{-x^2}$, whereas the natural tail of the
nuclear density is exponential, $e^{-x}$. Added flexibility is
desirable, and can be achieved by considering linear combinations
of harmonic oscillator wave functions~\cite{exps-1,exps-2,exps-3},
or different functional representations for the hole and particle
sides of the spectrum~\cite{cce-a1}. To this effect, in the second part of this paper, we will discuss the framework of calculating matrix
elements of the Argonne $v_{18}$ potential in a basis of single-particle
wave functions of arbitrary functional form. Here, the ability of using
the Moshinsky transformation brackets approach to perform the
separation of the center-of-mass and relative coordinates degrees of freedom is
rendered numerically unfeasible, as the required CPU time is of
order $N^4$ compared to the calculation involving Moshinsky transformation brackets.

For illustrative and testing purposes, we will apply this approach to
the case when the radial part of the single-particle wave functions, $R_{nl}(x)$ can be expanded out into harmonic-oscillator wave functions,
$\mathcal{HO}_{kl}(x)$, i.e.
\begin{equation}
   R_{nl}(x) \, = \, \sum_{k=1}^N \, A^k_{n l} \, \mathcal{HO}_{kl}(x)
   \>.
\label{rnl}
\end{equation}
This is convenient because we can still use a Gaussian quadrature technique
to perform the numerical computation of the radial integrals.
This also allows us to check of the reliability of our
matrix-elements calculation by comparing with matrix elements calculated
in a harmonic-oscillator basis using the procedures using the Moshinsky transformation brackets.

The paper is organized as follows: In Sec.~\ref{basics}, we review features of the Argonne $v_{18}$ potential, together with the definitions of the matrix elements to be discussed further. Basic aspects of the separation of the center-of-mass and
relative coordinates using the numerical implementation of the Talmi transformation via the Moshinsky transformation-brackets approach, together with the procedures required for the 7~types of operators part of the Argonne $v_{18}$ potential, are discussed in
Sec.~\ref{ops}. In Sec.~\ref{two} we present the general results used to calculate the
matrix elements of a translationally-invariant two-body interaction in an arbitrary single-particle basis. The implementation of this general approach for the particular case of the 7~operators in the Argonne $v_{18}$ potential is discussed in Sec.~\ref{app:jjme_app}. This calculation is carried out in a \emph{jj}-coupling scheme.
The part of the calculation involving the isospin part of the interaction is identical in both frameworks. A brief review of this calculation is the object of Sec.~\ref{iso}.
Finally, in Sec.~\ref{discuss} we discuss numerical convergence aspects of the matrix elements calculation corresponding to a $^{16}$O-like model space of harmonic-oscillator single-particle wave functions, and compare with results obtained
using the two frameworks.
We also include an appendix on the calculation of large sets of abscissas
and weights for the evaluation of integrals using Gauss-Hermite
quadrature formulas, followed by an appendix summarizing results relevant to the calculation of reduced matrix elements used in Sec.~\ref{app:jjme_app}.


\section{Preliminaries}
\label{basics}

Consider the two-body approximation of the Hamiltonian describing a
system consisting of protons and neutrons
\begin{eqnarray}
   \mathbf{H}
   & = &
   \sum_i T_i
   \, + \, \sum_{i<j} \, V_{ij}
   \>.
\label{eq:Hamiltonian_0}
\end{eqnarray}
In the {\em second-quantization} representation this becomes
\begin{eqnarray}
   \mathbf{H}
   = &&
   \sum_{\alpha \beta}
       \mathbf{a}^{\dag}_{\alpha}
       \langle \alpha | T | \beta \rangle
       \mathbf{a}_{\beta}
   +
       \frac{1}{2} \sum_{\alpha \beta \gamma \delta}
       \mathbf{a}^{\dag}_{\alpha} \mathbf{a}^{\dag}_{\beta}
       \langle \alpha \beta | V_{2N} | \delta \gamma \rangle
       \mathbf{a}_{\gamma} \mathbf{a}_{\delta}
   \>,
\end{eqnarray}
where Greek letters label the single-particle states $| \alpha
\rangle = | n l s j m_j ; \frac {1} {2} m_\tau \rangle$, with $n \ge 1$, $s =
\frac {1} {2}$, $j = l \pm \frac {1} {2}$ and
$m_\tau = + \frac {1} {2} ( - \frac {1} {2})$ -- for a proton
(neutron). The parity of these states is ${(-1)}^l$. In this work we
will assume that the radial part of the single-particle wave
functions, $R_{nl}(x)$, is simply given by the radial part of the
harmonic oscillator wave functions $\mathcal{HO}_{kl}(x)$, subject
to the normalization conditions
\begin{equation}
   \int_0^\infty \, [x^2 \, \mathrm{d}x] \, R_{nl}(x) \, R_{n'l'}(x)
       \, = \,  \delta_{n \, n'} \delta_{l \, l'}
   \>.
\end{equation}
Here,  $x \, = \, r / b$, with $b$ the oscillator parameter. We note
that the single-particle radial functions $\mathcal{R}_{nl}(r)$ are
defined in terms of the the radial functions $R_{nl}(x)$ such that
they satisfy the normalization condition
\begin{equation}
   \int_0^\infty \, [r^2 \, \mathrm{d}r] \,
          \mathcal{R}_{nl}(r) \, \mathcal{R}_{n'l'}(r)
       \, = \,  \delta_{n \, n'} \delta_{l \, l'}
   \> .
\end{equation}
Thus, we have
\begin{equation}
   \mathcal{R}_{nl}(r) \, = \, \frac{1}{b^{3/2}} \, R_{nl}(x)
\label{xversusr}
   \> .
\end{equation}


\subsection{Argonne $v_{18}$ potential}

The Argonne $v_{18}$ potential~\cite{av18} is an updated version of
the nonrelativistic Argonne potential~\cite{av14} that fits both
{\it np} data and {\it pp} data, as well as low-energy {\it nn} data
scattering parameters and deuteron properties. The potential was fit
directly to the Nijmegen {\it NN} scattering database, which
contains 1787 {\it pp} and 2514 {\it np} data in the range 0-350
MeV, together with the $nn$~scattering length measured in
$d(\pi^-,\gamma)nn$ experiments and the deuteron binding energy. The
fit has a $\chi^2$ per datum of 1.09.

The strong interaction part of the Argonne $v_{18}$ potential is
projected into an operator format with 18~terms:

i)~a charge-independent part that has 14~operator components (as in
the older Argonne $v_{14}$~\cite{av14}), featuring 7~basic
operators,
\begin{eqnarray}
   1, \ \sigma_i \cdot \sigma_j, \ S_{ij}, \ \ell \cdot S, \
   \ell^2, \ \ell^2 \, (\sigma_i \cdot \sigma_j), \ (\ell \cdot S)^2
   \>,
\end{eqnarray}
and their charge-independent counterparts
\begin{eqnarray}
   &&
   \tau_i \cdot \tau_j, \ (\sigma_i \cdot \sigma_j) \, (\tau_i \cdot \tau_j), \
   S_{ij} \, (\tau_i \cdot \tau_j), \ (\ell \cdot S) \, (\tau_i \cdot \tau_j), \
   \nonumber \\ &&
   \ell^2 \, (\tau_i \cdot \tau_j), \ \ell^2 \, (\sigma_i \cdot \sigma_j) \, (\tau_i \cdot \tau_j), \
   (\ell \cdot S)^2 \, (\tau_i \cdot \tau_j)
   \>;
\end{eqnarray}

ii)~a charge-independence breaking part that has three
charge-dependent operators
\begin{eqnarray}
   T_{ij}, \ (\sigma_i\cdot\sigma_j) \, T_{ij}, \ S_{ij} \, T_{ij}
   \>,
\end{eqnarray}
where $T_{ij} = 3 \, \tau_{zi} \tau_{zj} - \tau_i \cdot \tau_j$ is
the isotensor operator, defined analogous to the $S_{ij}$ operator,
and iii)~one charge-asymmetric operator
\begin{eqnarray}
   \tau_{zi} + \tau_{zj}
   \>.
\end{eqnarray}
In principle, there could be more charge-independence breaking
terms, such as $(\ell \cdot S)\, T_{ij}$ or $S_{ij}\, (\tau_{zi} +
\tau_{zj})$, but the scattering data are not sufficiently precise to
identify them at present. The potential also includes a complete
electromagnetic potential, containing Coulomb, Darwin-Foldy, vacuum
polarization, and magnetic moment terms with finite-size effects.


\subsection{Two-body matrix elements}

Matrix elements of the two-body interaction can be specified either
in particle-particle (\emph{pp}) coupling or in particle-hole (\emph{ph}) coupling. Either set of matrix elements completely specify the
two-body interaction. We define \emph{ph}-coupled matrix elements
as
\begin{align}
  &
  \phme{V}{\lambda}
  =
  \!\!\!\! \sum_{m_1 m_2 m_3 m_4} \!\!\!\!
  \mmme{V}
\label{eq:phme_def}
  \\ \nonumber & \times \
  \sgn{j_3 - m_3} \, \cg{j_1}{m_1}{j_3}{-m_3}{\lambda}{\mu} \,
  \sgn{j_2 - m_2} \, \cg{j_4}{m_4}{j_2}{-m_2}{\lambda}{\mu} \,
  \>.
\end{align}
Correspondingly, the \emph{pp}-coupled matrix elements are
defined as
\begin{align}
  \ppme{V}{L}
  = &
  \!\!\!\! \sum_{m_1 m_2 m_3 m_4} \!\!\!\!\!\!
  \mmme{V}
\label{eq:ppme_def}
  \cg{j_1}{m_1}{j_2}{m_2}{L}{M}
  \cg{j_3}{m_3}{j_4}{m_4}{L}{M}
  \>.
\end{align}
In practice, the calculation of the two-body matrix elements using a
harmonic-oscillator single-particle basis set is carried out in
\emph{pp}~coupling. Subsequently, \emph{ph}-coupled matrix elements
are evaluated from their \emph{pp} counterpart, using the
relationship
\begin{align}
   &
   \phme{V}{\lambda}
\label{eq:convert_ppph}
   =
   \sum_{L} \, (-)^{j_3+j_4+L} \, (2L+1) \,
            \wsj{j_1}{j_3}{\lambda}{j_4}{j_2}{L} \,
   \ppme{V}{L}
   \>.
\end{align}
Conversely, we have
\begin{align}
   &
   \ppme{V}{L}
\label{eq:convert_phpp}
   =
   \sum_{\lambda} \, (-)^{j_3+j_4+L} \, (2\lambda+1) \,
            \wsj{j_1}{j_2}{L}{j_4}{j_3}{\lambda} \,
   \phme{V}{\lambda}
   \>.
\end{align}

It is useful to introduce here some notations we will be using throughout
the rest of this paper. First, we note that,
by definition, the scalar product of two tensor operators of rank
$k$, $U^{(k)}$ and $V^{(k)}$, is introduced as
\begin{eqnarray}
  \left ( U^{(k)} \odot V^{(k)} \right )
  =
\label{scalar_def}
  \sum_q \ \sgn{q} \ U^{(k)}_q \ V^{(k)}_{-q}
  \>.
\end{eqnarray}
In turn, the definition of the spherical tensor product of two tensor operators, $A^{(k_1)}$ and $B^{(k_2)}$, is
\begin{eqnarray}
  \left [ A^{(k_1)} \otimes B^{(k_2)} \right ]^{(k)}_q
  =
\label{tensor_def}
   \sum_{q_1 q_2} \, \cg{k_1}{q_1}{k_2}{q_2}{k}{q} \,
   A^{(k_1)}_{q_1} B^{(k_2)}_{q_2}
  \>.
\end{eqnarray}

Second, we note that the angular degrees of freedom will be described in terms of the
\emph{unnormalized} spherical harmonics, defined in terms of their
\emph{normalized} counterparts such that (\cite{Edmonds},
Eq.~2.5.31)
\begin{equation}
   \ckq{k}{q}{r} \, = \, \frac{\sqrt{4 \pi}}{\hat k} \,  \Ykq{k}{q}{r}
   \>.
\end{equation}
The \emph{unnormalized} spherical harmonics satisfy the
orthogonality conditions
\begin{eqnarray}
   \int \mathrm{d}\Omega \ \ckq{k_1}{q_1}{r}{}^* \, \ckq{k_2}{q_2}{r}
   =
   \frac{4\pi}{2k_1+1} \, \delta_{k_1 \, k_2} \, \delta_{q_1 \, q_2}
   \>.
\label{ckortho}
\end{eqnarray}
For completeness, we list here useful properties of the
\emph{unnormalized}
spherical harmonics:\\
i)~$\ck{0}{r} \, = \, 1$.\\
   ii)~$C^{(k)\star}_q(\hat r) \, = \, (-)^{q} \, \ckq{k}{-q}{r}$.\\
   iii) $\left ( \ck{k}{r} \odot \ck{k}{p} \right ) \, = \, P_k(\cos \omega )$,
         where $\omega$ is the angle between $\hat{r}$ and
         $\hat{p}$.\\
   iv) $\left [ \ck{k_1}{r} \otimes \ck{k_2}{r} \right ]^{(k)} =
          \cg{k_1}{0}{k_2}{0}{k}{0} \, \ck{k}{r}$.
\\
   v) $\ckq{k_1}{q_1}{r} \ \ckq{k_2}{q_2}{r} =
         \sum_{k q} \
                  \ckq{k}{q}{r} \
                  \cg{k_1}{0}{k_2}{0}{k}{0} $
        $         \cg{k_1}{q_1}{k_2}{q_2}{k}{q}$.

Finally, in the interest of brevity, throughout the remainder of this paper we will follow the conventions, and will be referring
directly to the equation numbers, found in Ref.~\cite{Edmonds}.


\section{Matrix elements calculation using harmonic-oscillator wave functions}
\label{ops}

Provided that the single-particle wave functions are
harmonic-oscillator wave functions, one can use Talmi transformation~\cite{Talmi}
via the Moshinsky transformation brackets~\cite{brackets}
to exactly factorize a product of two
single-particle wave functions into a part which depends only on the
center-of-mass degrees of freedom, and a part which depends only on
the relative-motion degrees of freedom. For completeness, we will
review now the technical details of this procedure.


\subsection{Moshinsky transformation brackets}

Consider a two-particle system in a harmonic oscillator potential.
We shall characterize the two particles by their coordinates and
quantum numbers. For the purpose of our discussion, we introduce two
system of coordinates:
\begin{itemize}
   \item \emph{laboratory frame}, where the two particles are described by their
         coordinates with respect to the center of the potential well,
         $\vec r_1$ and $\vec r_2$, and corresponding radial, $n_1$ and $n_2$,
         and orbital quantum numbers, $l_1$ and $l_2$.
   \item \emph{center-of-mass frame}, where the system is characterized
         by the relative coordinate $\vec r$ and the coordinate $\vec R$ of the
         center of mass of the two particles, defined as
         \begin{equation}
            \vec r = \frac{1}{\sqrt{2}} \ ( \vec r_1 - \vec r_2 )
            \>,
            \quad
            \vec R = \frac{1}{\sqrt{2}} \ ( \vec r_1 + \vec r_2 )
            \>;
         \end{equation}
         the radial and orbital quantum numbers $n,l$ will correspond to the
         relative motion, and $N,L$ to that of the center of mass.
\end{itemize}
The eigenkets in the two coordinate systems may be written as
follows:
\begin{itemize}
   \item \emph{laboratory frame}
       \begin{equation}
          | n_1 n_2 \: (l_1 l_2) \, \lambda \mu \rangle
          = \!\!\!\!
          \sum_{m_1 m_2} \cg{l_1}{m_1}{l_2}{m_2}{\lambda}{\mu}
                             | n_1 l_1 m_1 \rangle  \ | n_2 l_2 m_2 \rangle
          \>;
       \label{eq:SL_2bwf}
       \end{equation}
   \item \emph{center-of-mass frame}
       \begin{equation}
          | n N \: (l L) \, \lambda \mu \rangle
          = \!\!
          \sum_{m M}  \cg{l}{m}{L}{M}{\lambda}{\mu}
                         | n l m \rangle  \ | N L M \rangle
          \>.
       \label{eq:CM_2bwf}
       \end{equation}
\end{itemize}
The Moshinsky transformation brackets are defined as the expansion coefficients of
the eigenket~(\ref{eq:SL_2bwf}) in a series of
eigenkets~(\ref{eq:CM_2bwf}). We have~\cite{brackets}
\begin{equation}
   | n_1 n_2 \: (l_1 l_2) \, \lambda \mu \rangle
   = \!\!\!
   \sum_{n l N L}
     \mc{n}{l}{N}{L}{n_1}{l_1}{n_2}{l_2}{\lambda}
     | n N \: (l L) \, \lambda \mu \rangle
   \>.
\end{equation}
This transformation is independent of the magnetic quantum number
$\mu$, and the transformation bracket vanishes for all combinations
of its parameters which do not satisfy the total angular momentum
\begin{equation}
   \vec{\lambda} \ = \ \vec \ell_1 + \vec \ell_2
                 \ = \ \vec \ell + \vec L
   \>,
\end{equation}
and energy
\begin{eqnarray}
   E
   & \propto &
   ( 2n_1 + l_1 + 3/2 ) \ + \ ( 2n_2 + l_2 + 3/2 )
   \nonumber \\
   & = &
   ( 2n + l + 3/2 ) \ + \ ( 2N + L + 3/2 )
   \>.
\label{eq:energy}
\end{eqnarray}
conservation laws. Therefore, the transformation bracket vanishes
for all combinations of its parameters which do not satisfy the
energy condition~(\ref{eq:energy}), and any summations over
$\lambda$ will be restricted by the corresponding triangle
relations:
\begin{eqnarray}
    | l_1 - l_2 | \leq \lambda \leq l_1 + l_2
    \>,
    \nonumber \\
    | l - L | \leq \lambda \leq l + L
    \>.
\end{eqnarray}

We calculate the two-body matrix element in a harmonic oscillator
single-particle basis using the Moshinsky transformation brackets, and obtain
\begin{align}
   &
   \langle n_1 n_2 \,
   {(l_1 {\textstyle\frac{1}{2}})}_{j_1} {(l_2 {\textstyle\frac{1}{2}})}_{j_2} ; \, J M_J
   \; | \; V \; | \; n_3 n_4 \,
   {(l_3 {\textstyle\frac{1}{2}})}_{j_3} {(l_4 {\textstyle\frac{1}{2}})}_{j_4} ; \, J M_J
   \rangle
   \nonumber \\ &
   =
   \hat j_1 \, \hat j_2 \
   \hat j_3 \, \hat j_4 \
   \!\!\!\!\!\!\!\! \sum_{n l N L\, \lambda ; n' l' N' L'\, \lambda'} \!\!\!\!\!\!\!\!
   \hat \lambda \, \hat \lambda' \
   \mc{n_1}{l_1}{n_2}{l_2}{n'}{l'}{N'}{L'}{\lambda '}
   \mc{n_3}{l_3}{n_4}{l_4}{n}{l}{N}{L}{\lambda}
   \nonumber \\ & \qquad \times \
   \sum_{S,S'=0}^1 \
   \hat S \, \hat S' \
   \wnj{l_1}{\frac{1}{2}}{j_1}{l_2}{\frac{1}{2}}{j_2}{\lambda'}{S'}{J} \
   \wnj{l_3}{\frac{1}{2}}{j_3}{l_4}{\frac{1}{2}}{j_4}{\lambda}{S}{J} \
   \\ \nonumber & \qquad \qquad \qquad \times \
   \langle n' N' \,
   (l' L')_{\lambda'} \, ({\textstyle\frac{1}{2}} \, {\textstyle\frac{1}{2}})_{S'} \, ; \, J M_J
   \; | \; V \; | \; n N \,
   (l L)_\lambda \, ({\textstyle\frac{1}{2}} \, {\textstyle\frac{1}{2}})_S \, ; \, J M_J
   \rangle
   \>.
\end{align}
For a particular choice of the potential, $V$, all we have to do is to
calculate the matrix element
\begin{equation}
   \langle n' N' \:
   (l' L')_{\lambda'} \, ({\textstyle\frac{1}{2}} \, {\textstyle\frac{1}{2}})_{S'} \, ; \, J M_J
   \; | \; V \; | \; n N \:
   (l L)_\lambda \, ({\textstyle\frac{1}{2}} \, {\textstyle\frac{1}{2}})_S \, ; \, J M_J
   \rangle
   \>.
\end{equation}


\subsection{Central interaction.}

For the case of the central interaction, the potential depends only
on the magnitude of the vector $\vec{r}$, and not on its angular
degrees of freedom. We apply the Wigner-Eckart theorem for the case
of a zero-rank tensor (\cite{Edmonds}, Eq.~5.4.1a)
\begin{eqnarray}
   &&
   \langle n' N' \:
   {(l' L')}_{\lambda'} {({\textstyle\frac{1}{2}} \, {\textstyle\frac{1}{2}})}_{S'} ; \: J M_J
   \; | \; V_c \; | \; n N \:
   {(l L)}_{\lambda} {({\textstyle\frac{1}{2}} \, {\textstyle\frac{1}{2}})}_{S} ; \: J M_J
   \rangle
\label{eq:c_medef}
   \\ \nonumber && =
   \frac{1}{\hat{J}} \
   \langle n' N' \:
   {(l' L')}_{\lambda'} {({\textstyle\frac{1}{2}} \, {\textstyle\frac{1}{2}})}_{S'} ; \: J M_J
   \; \| \; V_c \; \| \; n N \:
   {(l L)}_{\lambda} {({\textstyle\frac{1}{2}} \, {\textstyle\frac{1}{2}})}_{S} ; \: J M_J
   \rangle
   \>,
\end{eqnarray}
where we have introduced the notation $\hat J = \sqrt{2J+1}$. The
reduced matrix element in Eq.~(\ref{eq:c_medef}) is evaluated using
(\cite{Edmonds}, Eq.~7.1.7). We have
\begin{align}
   &
   \langle n' N' \:
   {(l' L')}_{\lambda'} {({\textstyle\frac{1}{2}} \, {\textstyle\frac{1}{2}})}_{S'} ; \: J M_J
   \; \| \; V_c \; \| \; n N \:
   {(l L)}_{\lambda} {({\textstyle\frac{1}{2}} \, {\textstyle\frac{1}{2}})}_{S} ; \: J M_J
   \rangle
\label{eq:c_redme1}
   \\ \nonumber & =
   \delta_{S S'} \ (-)^{\lambda'+S+J+0} \ (2J+1) \
   \wsj{\lambda'}{J}{S}{J}{\lambda}{0}
   \langle n' N' \: {(l' L')}_{\lambda'}
   \; \| \; V_c \; \| \; n N \: {(l L)}_{\lambda} \rangle
   \>.
\end{align}
We use (\cite{Edmonds}, Eq.~7.1.7) one more time for the reduced
matrix element
\begin{align}
   &
   \langle n' N' \: {(l' L')}_{\lambda'}
   \; \| \; V_c \; \| \; n N \: {(l L)}_{\lambda} \rangle
\label{eq:c_redme2}
   \\ \nonumber & =
   \delta_{N N'} \delta_{L L'}
   (-)^{l'+L+\lambda+0}
   \hat \lambda \, \hat \lambda'
   \wsj{l'}{\lambda'}{L}{\lambda}{l}{0}
   \langle n' \: l' \; \| \; V_c \; \| \; n \: l \rangle
   \>.
\end{align}
The Wigner $6j$ symbols in
Eqs.~(\ref{eq:c_redme1},\ref{eq:c_redme2}) are calculated using
(\cite{Edmonds}, Eq.~6.3.2).
From the Wigner-Eckart theorem (\cite{Edmonds},
Eq.~5.4.1a) we have:
\begin{eqnarray}
   \langle n' \: l' m_l \;  | \; V_c \;  | \; n \: l  m_l \rangle
   & = &
   \delta_{l l'} \ \frac{1}{\hat l} \
   \langle n' \: l' \; \| \; V_c \; \| \; n \: l \rangle
   \label{eq:c_merad}
   \\ \nonumber & = &
   \delta_{l l'} \ {\cal RM}[V_c](n' l \: ; \: n l)
   \>,
\end{eqnarray}
where we have introduced the notation
\begin{eqnarray}
   {\cal RM}[V](n' l' \: ; \: n l)
\label{radmat}
   & = &
   \! \int_0 ^{\infty} \! [ r^2 \, {\rm d}r ] \:
   \mathcal{HO}_{n' l'}(r) \, V(r) \, \mathcal{HO}_{n l}(r)
   \> .
\end{eqnarray}
Therefore, Eq.~(\ref{eq:c_medef}) becomes
\begin{eqnarray}
   &&
   \langle n' N' \:
   {(l' L')}_{\lambda'} {({\textstyle\frac{1}{2}} \, {\textstyle\frac{1}{2}})}_{S'} ; \: J M_J
   \; | \; V_c \; | \; n N \:
   {(l L)}_{\lambda} {({\textstyle\frac{1}{2}} \, {\textstyle\frac{1}{2}})}_{S} ; \: J M_J
   \rangle
\label{eq:mosh_dr}
   \\ \nonumber
   && =
   \delta_{S S'} \delta_{N N'} \delta_{L L'}
   \delta_{l l'} \delta_{\lambda \lambda'} \
   {\cal RM}[V_c](n' l \: ; \: n l)
   \>.
\end{eqnarray}


\subsection{Spin-spin interaction}

It is convenient to carry out the calculation of the spin-spin
interaction matrix element by using the \emph{m}-representation
approach outlined above. We have:
\begin{eqnarray}
   &&
   \langle n' N' \:
   {(l' L')}_{\lambda'} {({\textstyle\frac{1}{2}} \, {\textstyle\frac{1}{2}})}_{S'} ; \: J M_J
   \; | \; V_s \: \sigma_1 \cdot \sigma_2 \; | \; n N \:
   {(l L)}_{\lambda} {({\textstyle\frac{1}{2}} \, {\textstyle\frac{1}{2}})}_{S} ; \: J M_J
   \rangle
   \\ \nonumber && = \
   \sum_{m_{l'} M' m_{S'} \mu'} \!\!
   \cg{l'}{m_{l'}}{L'}{M'}{\lambda'}{\mu'}
   \cg{\lambda'}{\mu'}{S'}{m_{S'}}{J}{M_J}
   \\ \nonumber && \qquad \times \
   \sum_{m_l M m_S \mu} \
   \cg{l}{m_l}{L}{M}{\lambda}{\mu}
   \cg{\lambda}{\mu}{S}{m_S}{J}{M_J}
   \\ \nonumber && \qquad \quad \times \,
   \langle n' N' \: l' m_{l'} \: L' M' \: S' m_{S'}
   \; | \; V_s \, \sigma_1 \cdot \sigma_2 \; | \; n N \:
   n N \: l m_l \: L M \: S m_S
   \rangle
   \, ,
\end{eqnarray}
where the matrix element in the last equation is given by
\begin{eqnarray}
   &&
   \langle n' N' \: l' m_{l'} \: L' M' \: S' m_{S'}
   \; | \; V_s \, \sigma_1 \cdot \sigma_2 \; | \; n N \:
   n N \: l m_l \: L M \: S m_S
   \rangle
   \\ \nonumber && = \
   \delta_{N N'} \delta_{L L'} \delta_{M M'}
   \delta_{l l'} \delta_{m_l m_{l'}} \
   \langle n' l \; | \; V_s \; | \; n l \rangle \
   \langle ({\textstyle\frac{1}{2}} {\textstyle\frac{1}{2}})_{S'}
   \; | \; \sigma_1 \cdot \sigma_2 \; | \;
   ({\textstyle\frac{1}{2}} {\textstyle\frac{1}{2}})_{S} \rangle
   \>.
\end{eqnarray}
The spin-dependent factor is obtained as
\begin{eqnarray}
   \langle ({\textstyle\frac{1}{2}} {\textstyle\frac{1}{2}})_{S'}
   \; | \; \sigma_1 \cdot \sigma_2 \; | \;
   ({\textstyle\frac{1}{2}} {\textstyle\frac{1}{2}})_{S} \rangle
   &&
   \ = \
   2 \ \left [ S(S+1) - \frac{3}{2} \right ] \ \delta_{S S'}
   \>,
\end{eqnarray}
where we used the fact that $\hat{s}_i = \frac{1}{2}
\hat{\sigma}_i$. Therefore, we obtain
\begin{eqnarray}
   &&
   \langle n' N' \:
   {(l' L')}_{\lambda'} {({\textstyle\frac{1}{2}} \, {\textstyle\frac{1}{2}})}_{S'} ; \: J M_J
   \; | \; V_s \: \sigma_1 \cdot \sigma_2 \; | \; n N \:
   {(l L)}_{\lambda} {({\textstyle\frac{1}{2}} \, {\textstyle\frac{1}{2}})}_{S} ; \: J M_J
   \rangle
\label{eq:mosh_ss}
   \\ \nonumber && = \
   \delta_{N N'} \delta_{L L'} \delta_{M M'}
   \delta_{l l'} \delta_{S S'}
   \delta_{\lambda \lambda'} \
   2 \left [ S(S+1) - \frac{3}{2} \right ] \
   {\cal RM}[V_s](n' l \: ; \: n l)
   \>.
\end{eqnarray}


\subsection{Tensor interaction}

We begin the calculation of the tensor interaction matrix element by
using (\cite{Edmonds}, Eq.~7.1.6)
\begin{eqnarray}
   &&
   \langle n' N' \:
   {(l' L')}_{\lambda'} {({\textstyle\frac{1}{2}} \, {\textstyle\frac{1}{2}})}_{S'} ; \: J M_J
   \; | \; V_t \, S_{12} \; | \; n N \:
   {(l L)}_{\lambda} {({\textstyle\frac{1}{2}} \, {\textstyle\frac{1}{2}})}_{S} ; \: J M_J
   \rangle
\label{eq:te_me1}
   \\ \nonumber &&
   =
   (-)^{\lambda+S'+J} \
   \wsj{J}{S'}{\lambda'}{2}{\lambda}{S} \
   \sqrt{6} \
   \sum_{n'' N''}
   \langle n' N' \: (l' L')_{\lambda'}
   \; \| \; V_t \ C^{(2)} \; \| \;
   n'' N'' \: (l L)_{\lambda} \rangle
   \\ \nonumber &&
   \qquad \qquad \times
   \langle n'' N'' \: ({\textstyle\frac{1}{2}} {\textstyle\frac{1}{2}})_{S'}
   \; \|
   [ \sigma_1 \otimes \sigma_2 ]^{(2)}
   \; \| \;
   n N \: ({\textstyle\frac{1}{2}} {\textstyle\frac{1}{2}})_{S} \rangle
   \, ,
\end{eqnarray}
where
     \begin{eqnarray}
        S_{12} && \ = \
        3 \, (\sigma_1 \cdot \hat r) \ (\sigma_2 \cdot \hat r)
        \ - \ \sigma_1 \cdot \sigma_2
\label{S12_def}
        \ = \
        \sqrt{6} \
        \Bigl ( C^{(2)} (\hat{r}) \odot
                [ \sigma_1 \otimes \sigma_2 ]^{(2)}
        \Bigr )
        \>.
    \end{eqnarray}
The reduced matrix elements in Eq.~(\ref{eq:te_me1}) are evaluated
as follows:\\
\noindent (i) We use (\cite{Edmonds}, Eq.~7.1.5) -- for $S (S') =
0,1$,  to obtain
         \begin{align}
            \langle ({\textstyle\frac{1}{2}} {\textstyle\frac{1}{2}})_{S'}
            \; \|
            [ \sigma_1 \otimes \sigma_2 ]^{(2)}
            \| \;
            ({\textstyle\frac{1}{2}} {\textstyle\frac{1}{2}})_{S} \rangle
            =
            \sqrt{5} \, \hat S \hat S' \
            \wnj{\frac{1}{2}}{\frac{1}{2}}{1}
                {\frac{1}{2}}{\frac{1}{2}}{1}
                {S'}{S}{2}
           \langle {\textstyle\frac{1}{2}} \; \| \; \sigma_1 \; \| \;
           {\textstyle\frac{1}{2}} \rangle
           \langle {\textstyle\frac{1}{2}} \; \| \; \sigma_2 \; \| \;
           {\textstyle\frac{1}{2}} \rangle
         \end{align}
   with $
            \langle {\textstyle\frac{1}{2}} \; \| \; \sigma_i \; \| \;
            {\textstyle\frac{1}{2}} \rangle
            =
            \sqrt{6}
$~~(see Ref.~\cite{Edmonds}, Eq.~5.4.4).

\noindent (ii) Next, we use (\cite{Edmonds}, Eq.~7.1.7)  to obtain
         \begin{eqnarray}
            &&
            \langle n' N' \: (l' L')_{\lambda'}
            \; \| \; V_t \ C^{(2)} \; \| \;
            n N \: (l L)_{\lambda} \rangle
            \\ \nonumber && = \
            \delta_{N N'} \delta_{L L'}
            (-)^{l'+L+\lambda+2} \
            \hat \lambda \hat \lambda'
            \wsj{l'}{\lambda'}{L}{\lambda}{l}{2}
            \langle n' \: l'
            \; \| \; V_t \ C^{(2)} \; \| \;
            n \: l \rangle \>,
         \end{eqnarray}

\noindent (iii) We also have
         \begin{eqnarray}
            \langle n' \: l'
            \; \| \; V_t \ C^{(2)} \; \| \;
            n \: l \rangle && =
            \langle l'
            \; \| \; C^{(2)} \; \| \;
            l \rangle \
            {\cal RM}[V_t](n' l' \: ; \: n l)
            \>,
         \end{eqnarray}
         where
         $
            \langle l'
            \; \| \; C^{(k)} \; \| \;
            l \rangle
         $ is given by Eq.~(\ref{ck_red}).

We obtain
\begin{eqnarray}
   &&
   \langle n' N' \:
   {(l' L')}_{\lambda'} {({\textstyle\frac{1}{2}} \, {\textstyle\frac{1}{2}})}_{S'} ; \: J M_J
   \; | \; V_t \, S_{12} \; | \; n N \:
   {(l L)}_{\lambda} {({\textstyle\frac{1}{2}} \, {\textstyle\frac{1}{2}})}_{S} ; \: J M_J
   \rangle
\label{eq:mosh_te}
   \\ \nonumber && =
   \delta_{N N'} \delta_{L L'}
   \delta_{S S'} \delta_{S 1} \
   2 \, \sqrt{30} \
   (-)^{l'+L+J+1} \
   \hat \lambda \hat \lambda' \hat l \
   \cg{l}{0}{2}{0}{l'}{0} \
   \\ \nonumber && \qquad \times \
   \wsj{J}{S}{\lambda'}{2}{\lambda}{S}
   \wsj{l'}{\lambda'}{L}{\lambda}{l}{2}
   {\cal RM}[V_t](n' l' \: ; \: n l) \>.
\end{eqnarray}


\subsection{Spin-orbit interaction}

Similarly to the case of the tensor interaction, we first use
(\cite{Edmonds}, Eq.~7.1.6)
\begin{align}
   &
   \langle n' N' \:
   {(l' L')}_{\lambda'} {({\textstyle\frac{1}{2}} \, {\textstyle\frac{1}{2}})}_{S'} ; \: J M_J
   \; | \; V_{ls} \, \bigl ( \mathbf{\ell} \cdot \mathbf{S} \bigr ) \; | \; n N \:
   {(l L)}_{\lambda} {({\textstyle\frac{1}{2}} \, {\textstyle\frac{1}{2}})}_{S} ; \: J M_J
   \rangle
   \\ & \nonumber
   =
   (-)^{\lambda+S'+J}
   \wsj{J}{S'}{\lambda'}{1}{\lambda}{S}
   \\ & \nonumber \qquad \times \,
   \sum_{n'' N''}
   \langle n' N' \: (l' L')_{\lambda'}
   \; \| \; V_{ls} \, \mathbf{\ell} \; \| \;
   n'' N'' \: (l L)_{\lambda} \rangle
   \langle n'' N'' \: ({\textstyle\frac{1}{2}} {\textstyle\frac{1}{2}})_{S'}
   \; \| \mathbf{S} \; \| \;
   n N \: ({\textstyle\frac{1}{2}} {\textstyle\frac{1}{2}})_{S} \rangle
   \>,
\end{align}
with the reduced matrix elements calculated as:\\
\noindent (i) We use (\cite{Edmonds}, Eq.~5.4.3) -- for $S = 0,1$,
to obtain
         \begin{eqnarray}
            \langle ({\textstyle\frac{1}{2}} {\textstyle\frac{1}{2}})_{S'}
            \; \| \mathbf{S} \; \| \;
            ({\textstyle\frac{1}{2}} {\textstyle\frac{1}{2}})_{S} \rangle &&
            \ = \
\label{S1S}
            \delta_{S S'} \sqrt{S(S+1)(2S+1)}
            \ = \
            \sqrt{6} \ \delta_{S 1} \delta_{S S'} \>.
         \end{eqnarray}
\noindent (ii) Next, we use (\cite{Edmonds}, Eq.~7.1.7) to obtain
         \begin{eqnarray}
            &&
            \langle n' N' \: (l' L')_{\lambda'}
            \; \| \; V_{ls} \, \mathbf{\ell} \; \| \;
            n N \: (l L)_{\lambda} \rangle
            \\ \nonumber &&
            = \
            \delta_{N N'} \delta_{L L'} \
            (-)^{l'+L+\lambda+1} \
            \hat \lambda \hat \lambda' \
            \wsj{l'}{\lambda'}{L}{\lambda}{l}{1}
            \langle n' \: l'
            \; \| \; V_{ls} \, \mathbf{\ell} \; \| \;
            n \: l \rangle
            \>.
         \end{eqnarray}
\noindent (iii) Finally, we use~$\langle l
            \; \| \; \mathbf{\ell} \; \| \;
            l' \rangle = \delta_{l l'} \
            \hat l \ \sqrt{l(l+1)}$~~(\cite{Edmonds}, Eq.~5.4.3) to
obtain
         \begin{eqnarray}
            \langle n' \: l'
            \; \| \; V_{ls} \, \mathbf{\ell} \; \| \;
            n \: l \rangle
            =
            \delta_{l l'} \ \hat l \ \sqrt{l(l+1)} \
            {\cal RM}[V_{ls}](n' l \: ; \: n l)
            \>.
         \end{eqnarray}
Collecting terms, we find
\begin{eqnarray}
   \langle n' N' \:
   {(l' L')}_{\lambda'} {({\textstyle\frac{1}{2}} \, {\textstyle\frac{1}{2}})}_{S'} ; \: J M_J
   \; | \; V_{ls} \, \bigl ( \mathbf{\ell} \cdot \mathbf{S} \bigr ) \; | \; n N \:
   {(l L)}_{\lambda} {({\textstyle\frac{1}{2}} \, {\textstyle\frac{1}{2}})}_{S} ; \: J M_J
   \rangle
\label{eq:mosh_ls}
   \\ \nonumber =
   \delta_{N N'} \delta_{L L'}
   \delta_{l l'} \delta_{S S'} \delta_{S 1} \
   \sqrt{6} \
   (-)^{l+L+J} \
   \hat \lambda \hat \lambda' \hat l \
   \sqrt{l(l+1)} \
   \\ \nonumber \qquad \times \
   \wsj{J}{S}{\lambda'}{1}{\lambda}{S}
   \wsj{l}{\lambda'}{L}{\lambda}{l}{1}
   {\cal RM}[V_{ls}](n' l \: ; \: n l) \>.
\end{eqnarray}


\subsection{$\ell^2$ Interaction.}

The matrix element of the $\ell^2$ interaction can be easily calculated
when the matrix element is written in the \emph{m} representation,
i.e.
\begin{eqnarray}
   &&
   \langle n' N' \: l' m_{l'} \: L' M' \: S' m_{S'}
   \; | \; V_{l2} \, \ell^2 \; | \;
   n N \: l m_l \: L M \: S m_S
   \rangle
\label{eq:mosh_l2_m}
   \\ \nonumber && = \
   \delta_{N N'} \: \delta_{L L'} \: \delta_{M M'} \:
   \delta_{l l'} \: \delta_{m_l m_{l'}} \
   \delta_{S S'} \: \delta_{m_S m_{S'}} \ \ l(l+1) \ \
   {\cal RM}[V_{l2}](n' l \: ; \: n l)
   \>.
\end{eqnarray}
We use the transformation
\begin{eqnarray}
   &&
   \langle n' N' \:
   {(l' L')}_{\lambda'} {({\textstyle\frac{1}{2}} \, {\textstyle\frac{1}{2}})}_{S'} ; \: J M_J
   \; | \; V_{l2} \, \ell^2 \; | \; n N \:
   {(l L)}_{\lambda} {({\textstyle\frac{1}{2}} \, {\textstyle\frac{1}{2}})}_{S} ; \: J M_J
   \rangle
   \\ \nonumber && = \
   \sum_{m_{l'} M' m_{S'} \mu'}
   \cg{l'}{m_{l'}}{L'}{M'}{\lambda'}{\mu'}
   \cg{\lambda'}{\mu'}{S'}{m_{S'}}{J}{M_J}
   \\ \nonumber && \qquad \times \
   \sum_{m_l M m_S \mu} \
   \cg{l}{m_l}{L}{M}{\lambda}{\mu}
   \cg{\lambda}{\mu}{S}{m_S}{J}{M_J} \
   \\ \nonumber && \qquad \qquad \qquad \times \
   \langle n' N' \: l' m_{l'} \: L' M' \: S' m_{S'}
   \; | \; V_{l2} \, \ell^2 \; | \;
   n N \: l m_l \: L M \: S m_S
   \rangle
   \>,
\end{eqnarray}
substitute Eq.~(\ref{eq:mosh_l2_m}), and employ the orthogonality
properties of the Clebsch-Gordon coefficients, to obtain
\begin{eqnarray}
   &&
   \langle n' N' \:
   {(l' L')}_{\lambda'} {({\textstyle\frac{1}{2}} \, {\textstyle\frac{1}{2}})}_{S'} ; \: J M_J
   \; | \; V_{l2} \, \ell^2 \; | \; n N \:
   {(l L)}_{\lambda} {({\textstyle\frac{1}{2}} \, {\textstyle\frac{1}{2}})}_{S} ; \: J M_J
   \rangle
\label{eq:mosh_l2}
   \\ \nonumber && = \
   \delta_{\lambda \lambda'}
   \delta_{N N'} \delta_{L L'}
   \delta_{l l'} \delta_{S S'} \
   l(l+1) \
   {\cal RM}[V_{l2}] (n' l \: ; \: n l)
   \>.
\end{eqnarray}


\subsection{$\ell^2 \,\bigl ( \sigma_1 \cdot \sigma_2 \bigr )$ interaction}

The derivation of the matrix element for the $\ell^2 \, \bigl (
\sigma_1 \cdot \sigma_2 \bigr )$ interaction follows closely the
calculation of the matrix element corresponding to the $\ell^2$
interaction. We first perform a transformation to the \emph{m}
representation where the calculation of the matrix element is
particularly simple, i.e.
\begin{eqnarray}
   \langle n' N' \:
   {(l' L')}_{\lambda'} {({\textstyle\frac{1}{2}} \, {\textstyle\frac{1}{2}})}_{S'} ; \: J M_J
   \; | \; V_{l2s} \, \ell^2 \, \bigl ( \sigma_1 \cdot \sigma_2 \bigr ) \; | \; n N \:
   {(l L)}_{\lambda} {({\textstyle\frac{1}{2}} \, {\textstyle\frac{1}{2}})}_{S} ; \: J M_J
   \rangle
   \\ \nonumber = \
   \sum_{m_S \mu ; m_{S'} \mu'} \!\!\!\!
   \cg{\lambda'}{\mu'}{S'}{m_{S'}}{J}{M_J}
   \cg{\lambda}{\mu}{S}{m_S}{J}{M_J} \
   \\ \nonumber \qquad \qquad \times \
   \langle n' N' \: \lambda' \mu' \: S' m_{S'}
   \; | \; V_{l2s} \, \ell^2 \, \bigl ( \sigma_1 \cdot \sigma_2 \bigr ) \; | \;
   n N \: \lambda \mu \: S m_S
   \rangle
   \>,
\end{eqnarray}
which gives:
\begin{align}
   &
   \langle n' N' \:
   {(l' L')}_{\lambda'} {({\textstyle\frac{1}{2}} \, {\textstyle\frac{1}{2}})}_{S'} ; \: J M_J
   \; | \; V_{l2s} \, \ell^2 \, \bigl ( \sigma_1 \cdot \sigma_2 \bigr ) \; | \; n N \:
   {(l L)}_{\lambda} {({\textstyle\frac{1}{2}} \, {\textstyle\frac{1}{2}})}_{S} ; \: J M_J
   \rangle
\label{eq:mosh_l2s}
   \\ & \nonumber = \
   \delta_{\lambda \lambda'} \:
   \delta_{S S'} \:\delta_{N N'} \: \delta_{L L'} \:
   \delta_{l l'} \:
   {2} \Bigl [ S(S+1) - \frac{3}{2} \Bigr ] \ l(l+1) \
   {\cal RM}[V_{l2s}](n' l \: ; \: n l) \>.
\end{align}


\subsection{Quadrupole spin-orbit interaction}

For the purpose
of generalizing this matrix-element calculation to the case of an arbitrary set of
single-particle  wave functions, it is convenient to approach this
calculation by first noting that
\begin{eqnarray}
   (\mathbf{\ell} \cdot \mathbf{S})^2
   \ = \
   \sum_{j=0}^2 \ \hat j \
   \left [ \
   \left [ \mathbf{\ell} \otimes \mathbf{\ell} \right ]^{(j)}
   \otimes
   \left [ \mathbf{S} \otimes \mathbf{S} \right ]^{(j)} \
   \right ]^{(0)}
\label{ls2_def}
   \>.
\end{eqnarray}
A close inspection of the above equation shows that some of the
pieces of the quadrupole spin-orbit interaction may be incorporated
into the calculation of the other previous interactions involving
the relative orbital angular momentum operator, $\mathbf{\ell}$.\\

\noindent(i)~\emph{Case $j=0$.}
      \begin{eqnarray}
          \left [ \
          \left [ \mathbf{\ell} \otimes \mathbf{\ell} \right ]^{(0)}
          \otimes
          \left [ \mathbf{S} \otimes \mathbf{S} \right ]^{(0)} \
          \right ]^{(0)}
          = && \ \frac{1}{6} \ \mathbf{\ell}^2 \, \left ( 3 + \sigma_1\cdot\sigma_2 \right )
          \>,
      \end{eqnarray}
      where we have used
      ($\sigma_{i\,x}^2=\sigma_{i\,y}^2=\sigma_{i\,z}^2=1$)
      \begin{eqnarray}
        S^2
        = && \ \frac{1}{2} \ \left ( 3 + \sigma_1\cdot\sigma_2 \right )
        \>.
      \end{eqnarray}
      Therefore, we can introduce the modified radial amplitudes of the
      $\mathbf{\ell}^2$ and $\mathbf{\ell}^2 \, \bigl ( \sigma_1 \cdot \sigma_2 \bigr )$ interactions:
      \begin{eqnarray}
          \tilde{V}_{l2}   & = & \ V_{l2} \ + \ \frac{1}{2} \ V_{ls2}
\label{l2_mod}
          \>,
          \\
          \tilde{V}_{l2ss} & = & V_{l2ss} \ + \ \frac{1}{6} \ V_{ls2}
\label{l2s_mod}
          \>.
      \end{eqnarray}

\noindent(ii)~\emph{Case $j=1$.}
      \begin{eqnarray}
          \sqrt{3} \
          \Bigl [ \
          \bigl [ \mathbf{\ell} \otimes \mathbf{\ell} \bigr ]^{(1)}
          \otimes
          \bigl [ \mathbf{S} \otimes \mathbf{S} \bigr ]^{(1)} \
          \Bigr ]^{(0)}
          = && \
          - \ \frac{1}{2} \ \mathbf{\ell} \cdot \mathbf{S}
          \>,
      \end{eqnarray}
      where we have used the definition of the angular momentum quantum operator,
$
         \vec{J} \times \vec{J} = \mathrm{i} \, \vec J
$.
      Similarly, we introduce the modified radial amplitude for the spin-orbital
      interaction, as:
      \begin{equation}
          \tilde{V}_{ls} \  =  \ V_{ls} - \frac{1}{2} \ V_{ls2}
\label{ls_mod}
          \>.
      \end{equation}

\noindent(iii)~\emph{Case $j=2$.}
      \begin{eqnarray}
          \sqrt{5} \
          \left [ \
          \left [ \mathbf{\ell} \otimes \mathbf{\ell}\right ]^{(2)}
          \otimes
          \left [ \mathbf{S} \otimes \mathbf{S} \right ]^{(2)} \
          \right ]^{(0)}
\label{j2}
          \ = \
          \frac{\sqrt{5}}{2} \
          \left [ \,
          \left [ \mathbf{\ell} \otimes \mathbf{\ell}\right ]^{(2)}
          \otimes
          \left [ \sigma_1 \otimes \sigma_2 \right ]^{(2)} \,
          \right ]^{(0)}
          \>.
      \end{eqnarray}
We notice that the only component of the $\bigl ( \mathbf{\ell}
\cdot \mathbf{S} \bigr )^2$ interaction that we have not addressed
yet is the one corresponding to $j=2$.

To calculate the corresponding matrix element we first use
(\cite{Edmonds}, Eq.~7.1.6)
\begin{align}
   &
   \langle n' N' \:
   {(l' L')}_{\lambda'} {({\textstyle\frac{1}{2}} \, {\textstyle\frac{1}{2}})}_{S'} ; J M_J
   | V_{ls2}
   \hat j
   \left [
          \left [ \mathbf{\ell} \otimes \mathbf{\ell}\right ]^{(j)}
          \otimes
          \left [ \mathbf{S} \otimes \mathbf{S} \right ]^{(j)}
          \right ]^{(0)}
   | n N \:
   {(l L)}_{\lambda} {({\textstyle\frac{1}{2}}
    {\textstyle\frac{1}{2}})}_{S} ; J M_J
   \rangle
   \nonumber \\ &
   =
   (-)^{\lambda+S'+J+j}
   \wsj{J}{S'}{\lambda'}{j}{\lambda}{S}
   \sum_{n'' N''}
   \langle n' N' \: (l' L')_{\lambda'}
   \; \| \; V_{ls2} \, \left [ \mathbf{\ell} \otimes \mathbf{\ell}\right ]^{(j)} \; \| \;
   n'' N'' \: (l L)_{\lambda} \rangle
   \nonumber \\ & \quad \times \
   \langle n'' N'' \: ({\textstyle\frac{1}{2}} {\textstyle\frac{1}{2}})_{S'}
   \; \| \left [ \mathbf{S} \otimes \mathbf{S} \right ]^{(j)} \; \| \;
   n N \: ({\textstyle\frac{1}{2}} {\textstyle\frac{1}{2}})_{S} \rangle
   \>,
\end{align}
with the reduced matrix elements calculated as:\\

\noindent (i) We use (\cite{Edmonds}, Eq.~7.1.1), to obtain
         \begin{eqnarray}
            \langle ({\textstyle\frac{1}{2}} {\textstyle\frac{1}{2}})_{S'}
            \; \| \left [ \mathbf{S} \otimes \mathbf{S} \right ]^{(j)} \; \| \;
            ({\textstyle\frac{1}{2}} {\textstyle\frac{1}{2}})_{S} \rangle
            =
            \delta_{SS'} \delta_{S1} \
            6 \
            (-)^{j} \hat j
            \wsj{1}{1}{j}{1}{1}{1}
            \>.
         \end{eqnarray}

\noindent (ii) Next, we use (\cite{Edmonds}, Eq.~7.1.7) to obtain
         \begin{align}
            &
            \langle n' N' \: (l' L')_{\lambda'}
            \; \| \; V_{ls2} \, \left [ \mathbf{\ell} \otimes \mathbf{\ell}\right ]^{(j)} \; \| \;
            n N \: (l L)_{\lambda} \rangle
            \\ \nonumber & =
            \delta_{N N'} \delta_{L L'} \
            (-)^{l'+L+\lambda+j} \
            \hat \lambda \hat \lambda' \
            \wsj{l'}{\lambda'}{L}{\lambda}{l}{j}
            \langle n' \: l'
            \| V_{ls2} \, \left [ \mathbf{\ell} \otimes \mathbf{\ell}\right ]^{(j)} \|
            n \: l \rangle
            \>,
         \end{align}
         with (\cite{Edmonds}, Eq.~7.1.1)
         \begin{align}
            &
            \langle (n' l'
            \; \| V_{ls2} \, \left [ \mathbf{\ell} \otimes \mathbf{\ell}\right ]^{(j)} \; \| \;
            n l \rangle
            \nonumber \\ &
            =
            \delta_{ll'} \
            (-)^{j} \ \hat j \ \bigl [ l (l+1) (2l+1) \bigr ] \
            \wsj{1}{1}{j}{l}{l}{l}
            {\cal RM}[V_{ls2}](n' l \: ; \: n l)
            \>.
         \end{align}
Collecting terms, we find
\begin{align}
   &
   \langle n' N'
   {(l' L')}_{\lambda'} {({\textstyle\frac{1}{2}} {\textstyle\frac{1}{2}})}_{S'} ; J M_J
   | V_{ls2}
   \left [
          \left [ \mathbf{\ell} \otimes \mathbf{\ell}\right ]^{(j)}
          \otimes
          \left [ \mathbf{S} \otimes \mathbf{S} \right ]^{(j)}
   \right ]^{(0)}
   | n N
   {(l L)}_{\lambda} {({\textstyle\frac{1}{2}} {\textstyle\frac{1}{2}})}_{S} ; J M_J
   \rangle
   \nonumber \\ &
   =
\label{eq:mosh_ls2}
   \delta_{N N'} \delta_{L L'}
   \delta_{l l'} \delta_{S S'} \delta_{S 1} \
   6
   (-)^{l+L+J+1} \
            \hat j \hat \lambda \hat \lambda' \
            \bigl [ l (l+1) (2l+1) \bigr ]
   \\ \nonumber &
      \quad \times
            \wsj{1}{1}{j}{1}{1}{1}
            \wsj{1}{1}{j}{l}{l}{l}
   \wsj{J}{S'}{\lambda'}{j}{\lambda}{S}
   \wsj{l}{\lambda'}{L}{\lambda}{l}{j}
   {\cal RM}[V_{ls2}](n' l \: ; \: n l)
   \>.
\end{align}


\section{Two-body matrix elements calculation}
\label{two}

Unlike the case of harmonic-oscillator single-particle basis set,
the calculation (and storage) of the two-body matrix elements using
single-particle wave functions that can be represented as linear combinations of harmonic-oscillator
wave functions is performed more
efficiently in \emph{ph}~coupling.
The \emph{ph}-coupled matrix-element calculation is based on the
following two lemmas:

\textit{Lemma 1:
   Provided that the two-body potential can be factorized into parts
   depending only on the $\vec{r}_1$ or $\vec{r}_2$ coordinates, respectively,
   \begin{equation}
      V(\vec r_1, \vec r_2) \, = \,
            \left ( U^{(k)}(\vec r_1) \odot V^{(k)}(\vec r_2) \right )
      \>,
   \end{equation}
   then \emph{ph}-coupled matrix elements of the two-body interaction are given by
   \begin{align}
      &
      \phme{ \left ( U^{(k)}(\vec r_1) \odot V^{(k)}(\vec r_2) \right ) }{\lambda}
      \\ \nonumber &
      =
      \frac{\sgn{j_2+j_4+1}}{2\lambda+1} \,
            \sprme{1}{U^{(\lambda)}}{3} \,
                  \sprme{2}{V^{(\lambda)}}{4} \, \del{k}{\lambda}
      \>.
   \label{eq:lemma1}
   \end{align}
}
Using the definition of the scalar product of two tensor operators of rank
$k$, Eq.~(\ref{scalar_def}),
we calculate the matrix elements of the $U^{(k)}_q(\vec r_1)
V^{(k)}_{-q}(\vec r_2)$ operator in the \emph{m}-representation, using
the Wigner-Eckart theorem
\begin{align}
    \mmme{U^{(k)}_q(\vec r_1) V^{(k)}_{-q}(\vec r_2) }
    & = \,
        \frac{\sgn{j_3 - m_3}}{\hat k} \, \cg{j_1}{m_1}{j_3}{-m_3}{k}{q}
    \sprme{1}{U^{(k)}}{3} \,
        \nonumber \\ & \times \,
    \frac{\sgn{j_4 - m_4}}{\hat k} \, \cg{j_2}{m_2}{j_4}{-m_4}{k}{-q}
        \sprme{2}{V^{(k)}}{4}
    \>,
\end{align}
where we have also introduced the notation $\hat k = \sqrt{2k+1}$.
Then, Eq.~(\ref{eq:lemma1}) is obtained using the orthonormality of
the Clebsch-Gordon coefficients, together with the definition of
\emph{ph}-coupled matrix elements, Eq.~(\ref{eq:phme_def}).

We note the identity:
\begin{equation}
   \Bigl [ U^{(k)}(\vec r_1) \otimes V^{(k)}(\vec r_2) \Bigr ]^{(0)}
   = \frac{(-)^k}{\hat k} \,
   \Bigl ( U^{(k)}(\vec r_1) \odot V^{(k)}(\vec r_2) \Bigr )
   \>,
\end{equation}
where the spherical tensor product of two tensor operators, $A^{(k_1)}$ and $B^{(k_2)}$, was defined in Eq.~(\ref{tensor_def}).

\textit{Lemma 2:
   Provided that the spatial part of the two-body interaction has
   the form $V(r) \, \ck{k}{r}$, with $k$ a positive integer or zero,
   then the variables $\vec{r}_1$ and $\vec{r}_2$ can be separated in
   the sense that~\cite{horie}
   \begin{align}
      r^\alpha \, V(r) \, \ck{k}{r}
\label{lemma2}
      = \!\!
        \sum_{k_1,k_2} & \
            \mathrm{i}^{k_2-k_1-k} \frac{(2k_1+1)(2k_2+1)}{2k+1}
            \cg{k_1}{0}{k_2}{0}{k}{0} \,
      \\ \nonumber & \times \
        u^{(k_1 k_2 ; k)} (r_1,r_2)
        \left [ \ck{k_1}{r_1} \otimes \ck{k_2}{r_2} \right ]^{(k)}
\!\!      ,
   \end{align}
   where $\alpha$ is a positive integer exponent, and
   $\vec r = \vec r_1 - \vec r_2$ indicates the relative-motion coordinate. We have also introduced the notations
   \begin{equation}
      u^{(k_1 k_2 ; k \alpha)} (r_1,r_2) \, = \,
            \frac{2}{\pi} \, \intr{p} \ v_{\alpha \, \kappa}(p) \,
                                             j_{k_1}(p r_1) \,
                                             j_{k_2}(p r_2)
      \>,
   \end{equation}
   and
\begin{eqnarray}
   v_{\alpha \, \kappa}(p)
   \, = \,    \int_0^\infty \, [r^2 \, \mathrm{d}r] \, r^\alpha \, V(r) \, j_\kappa (p r)
\label{3_6}
   \> .
\end{eqnarray}
}

To prove lemma~2, we first introduce the asymmetric Fourier transform
of the operator $V(r) \, \ck{k}{r}$
\begin{equation}
   \tilde{v}^{(\alpha \, k)}(\vec{p}) \, = \,
      \frac{1}{(2\pi)^3} \, \intt{r} \ r^\alpha \, V(r) \ \ck{k}{r} \, e^{\mathrm{i} \vec{p}\, \cdot\vec{r}}
\label{eq:FT_VCk_def}
\end{equation}
and, conversely,
\begin{equation}
   r^\alpha \, V(r) \ \ck{k}{r} \, = \,
      \intt{p} \ \tilde{v}^{(k)}(\vec{p}) \, e^{-\mathrm{i} \vec{p}\, \cdot\vec{r}}
   \> .
\label{eq:invFT_VCk}
\end{equation}
We use the plane wave expansion in terms of \emph{unnormalized}
spherical harmonics,
\begin{eqnarray}
   e^{\mathrm{i} \vec{p}\, \cdot \vec{r}}
   \ = \
   \sum_{l} \, \mathrm{i}^l \, (2l+1) \, j_l (pr) \,
              \left ( \ck{l}{r} \odot \ck{l}{p} \right )
   \>,
\label{eq:plane_wave}
\end{eqnarray}
together with the orthogonality conditions~(\ref{ckortho}), and
carry out the angular part of the integral~(\ref{eq:FT_VCk_def}).
Then, $\tilde{v}^{(k)}(\vec{p})$ becomes
\begin{equation}
   \tilde{v}^{(\alpha \, k)}(\vec{p}) =
   \frac{(-\mathrm{i})^k}{2\pi^2} \ \ck{k}{p} \ v_{\alpha \, \kappa}(p)
   \>.
\label{eq:FT_VCk}
\end{equation}
From Eq.~(\ref{eq:invFT_VCk}), we can write
\begin{eqnarray}
   r^\alpha \, V(r) \ \ck{k}{r}
   \, = \,
      \frac{(-\mathrm{i})^k}{2\pi^2}
   \intt{p} \ v_{\alpha \, \kappa}(p) \ \ck{k}{p} \, e^{\mathrm{i} \vec{p}\, \cdot\vec{r}}
   \>.
\end{eqnarray}
The angular part of the last integral can be carried out explicitly
using the definition of the relative coordinate, $\vec{r} =
\vec{r}_1 - \vec{r}_2$, and by applying Eq.~(\ref{eq:plane_wave})
twice for $\exp(\mathrm{i}\, \vec{p}\cdot\vec{r}_1)$ and
$\exp(-\mathrm{i}\, \vec{p}\cdot\vec{r}_2)$, respectively. The
following integral is evaluated using the above properties of the
\emph{unnormalized} spherical harmonics:
\begin{align}
    \int \mathrm{d}\Omega_p & \ [\ckq{k_1}{q_1}{p}]^* \,
[\ckq{k_2}{q_2}{p}]^* \, \ckq{k}{q}{p}
   =
   \frac{4\pi}{2k+1} \,
      \cg{k_1}{0}{k_2}{0}{k}{0} \,
         \cg{k_1}{q_1}{k_2}{q_2}{k}{q}
   \>.
\end{align}
To finalize our proof, we use the definition of the spherical
tensor product of rank $k$ obtained from the tensor operators $\ck{k_1}{r_1}$ and $\ck{k_2}{r_2}$.


\section{Matrix elements calculation using arbitrary single-particle wave functions}
\label{app:jjme_app}

It is convenient to evaluate two-body matrix elements in \emph{ph}
angular momentum coupling using lemma~1. Then, according to lemma~2,
\emph{ph} matrix elements factorize in two parts, which depend on
the coordinates of the first and the second particle, respectively.
Hence, the appropriate angular-momentum coupling for each
single-particle wave function $| \alpha \rangle $ is provided in the
$(lsj)$ coupling, with the individual orbital angular
momentum~$l_\alpha$ and spin~$s_\alpha$ are coupled to a total
angular momentum~$j_\alpha$. For an operator $\mathcal{O}^{(k)}$ which
depends only on the orbital angular momentum
components, in the $(lsj)$-coupling scheme we have:\\

i) (see \cite{Edmonds}, 7.1.7)
      \begin{align}
        &
        \spjrme{1}{\mathcal{O}^{(k)}}{2}
        = \,
                \Ffac{1}{2}{k} \, \lropme{1}{\mathcal{O}^{(k)}}{2}
        \>,
      \end{align}
      with
      \begin{eqnarray}
        \Ffac{1}{2}{k} \, = \, \sgn{l_1+j_2+{\textstyle \frac{1}{2}}+k} \,
                \hat{j}_1 \hat{j}_2 \,
                        \wsj{l_1}{j_1}{\textstyle \frac{1}{2}}{j_2}{l_2}{k}
        \>.
      \end{eqnarray}

ii) (see \cite{Edmonds}, 7.1.5 and 5.4.4)
      \begin{align}
        &
        \spjrme{1}{\left [ \mathcal{O}^{(k_1)} \otimes \sigma \right ]^{(k_2)} }{2}
        = \,
                \Gfac{1}{2}{k_1}{k_2} \, \lropme{1}{\mathcal{O}^{(k_1)}}{2}
        \>,
      \end{align}
      with
      \begin{eqnarray}
        \Gfac{1}{2}{k_1}{k_2} \, = \, \sqrt{6} \, \hat{j}_1 \hat{j}_2 \hat{k}_2 \,
                \wnj{l_1}{l_2}{k_1}
            {\textstyle \frac{1}{2}}{\textstyle \frac{1}{2}}{1}
            {j_1}{j_2}{k_2}
        \>.
      \end{eqnarray}

We can now proceed to discussing the calculation of the two-body
matrix elements of the Argonne $v_{18}$ potential: We will first
evaluate the radial part of the matrix element, which can be
performed independently of the way we handle the angular and
spin/isospin degrees of freedom of the interaction. Next, we will
discuss the angular and spin part of the 7~basic operators
corresponding to the Argonne $v_{18}$ potential.


\subsection{Radial part of the two-body matrix element}
\label{seq:radial}

The radial part of the two-body matrix elements is defined as
\begin{align}
   R^{\alpha \, \kappa; \, k_1 k_2}_{n_1 l_1 \, n_2 l_2 \, ; \, n_3 l_3 \, n_4 l_4}
     = & \
     \frac{2}{\pi} \, \int_0^\infty \,
     [p^2 \, \mathrm{d}p] \, v_{\alpha \, \kappa}(p) \,
   \nonumber \\ & \times
   \int_0^\infty \, [r_1^2 \, \mathrm{d}r_1] \, \mathcal{R}_{n_1 l_1}(r_1) \,
       \Bigl [ \mathcal{O}^{(1)} \mathcal{R}_{n_3 l_3}(r_1) \Bigr ] \,
       j_{k_1}(p r_1) \,
   \nonumber \\ & \times
\label{ralphakappa}
   \int_0^\infty \, [r_2^2 \, \mathrm{d}r_2] \, \mathcal{R}_{n_2 l_2}(r_2) \,
       \Bigl [ \mathcal{O}^{(2)} \mathcal{R}_{n_4 l_4}(r_2) \Bigr ] \,
       j_{k_2}(p r_2)
   \> ,
\end{align}
where $v_{\alpha \, \kappa}(p)$ was defined in Eq.~(\ref{3_6}) and the operators
$\mathcal{O}^{(1,2)}$ are operators acting only on the radial part of the single-particle wave functions.
The functional form of the radial wave functions in
Eq.~(\ref{ralphakappa}) is not restricted to that of a
harmonic-oscillator wave function. Therefore, at this point we need
to carefully consider what is the most efficient numerical strategy
to computing~(\ref{ralphakappa}) for the particular radial form of
the wave functions under consideration. For illustrative (and
testing) purposes, in the following we will confine our discussion
to the case of a basesis of single-particle wave functions defined as linear combinations of harmonic-oscillator wave functions [see Eq.~(\ref{rnl})],
in which case some of the integrals in~(\ref{ralphakappa}) can be efficiently performed using Gauss-Hermite quadrature formulas.

Treating separately the integral
\begin{equation}
   \int_0^\infty \, [r_1^2 \, \mathrm{d}r_1] \, \mathcal{R}_{n_1 l_1}(r_1) \,
                    \Bigl [ \mathcal{O}^{(1)} \mathcal{R}_{n_3 l_3}(r_1) \Bigr ] \,
                                             j_{k_1}(p r_1)
   \> ,
\label{eq:int1_2b}
\end{equation}
we first change variables such that {$x_i' = r_i\ (\sqrt 2/b)$ with
$q'$~$=$~$p \ (b/\sqrt{2})$}, and we recall that (see discussion surrounding Eq.~(\ref{xversusr})
\begin{equation}
   R_{nl}({\scriptstyle \frac{1}{\sqrt{2}}} \, {x'})
   \, = \, b^{3/2} \, R_{nl}(r)
   \>.
\end{equation}
Next, we perform the harmonic-oscillator
expansion
\begin{eqnarray}
   R_{n_1 l_1}({\scriptstyle \frac{1}{\sqrt{2}}} \, {x'}_1) \,
&      \Bigl [
      \mathcal{O}^{(1)} R_{n_3 l_3}({\scriptstyle \frac{1}{\sqrt{2}}} \, {x'}_1)
   \Bigr ]
   \, = \,    \sum_n \, A^{n k_1}_{n_1 l_1 \, n_3 l_3} \, \mathcal{HO}_{n k_1}({x'}_1)
   \>.
\end{eqnarray}
Therefore, the integral (\ref{eq:int1_2b}) becomes
\begin{eqnarray}
   \frac{1}{( \sqrt{2} )^3} \,    \sum_n \, A^{n k_1}_{n_1 l_1 \, n_3 l_3} \,
      \left [ \sqrt{\frac{\pi}{2}} \, \widetilde{\mathcal{HO}}_{n k_1}(q')
   \right ]
   \>,
\label{eq:intr1_2b}
\end{eqnarray}
where we have introduced the notation, $\widetilde{\mathcal{HO}}_{n\ell}(q) = (-)^n \, \mathcal{HO}_{n\ell}(q)$, which denotes
the Fourier transform of the harmonic-oscillator wave function,
$\mathcal{HO}_{n\ell}$, and the coefficients $A^{n k_1}_{n_1 l_1 \,
n_3 l_3}$ are calculated as
\begin{eqnarray}
   A^{n k_1}_{n_1 l_1 \, n_3 l_3} \, =
      \int_0^\infty & [{x'}_1^2 \, \mathrm{d}x'_1] \,
                         R_{n_1 l_1}({\scriptstyle \frac{1}{\sqrt{2}}} \, {x'}_1) \,
\label{3_10}
   \Bigl [
                   \mathcal{O}^{(1)} R_{n_3 l_3}({\scriptstyle \frac{1}{\sqrt{2}} \, {x'}_1})
                   \Bigr ] \,
                                      \mathcal{HO}_{n k_1}({x'}_1)
   \> .
\end{eqnarray}
This integral is conveniently performed using Gaussian quadratures.
Similarly, we have:
\begin{align}
   &
   \int_0^\infty \!\! [r_2^2 \mathrm{d}r_2] \mathcal{R}_{n_2 l_2}(r_2)
                    \Bigl [ \!\mathcal{O}^{(2)} \mathcal{R}_{n_4 l_4}(r_2) \Bigr ]
                                             j_{k_2}(p r_2)
   \nonumber \\ &
   =
   \frac{1}{( \sqrt{2} )^3}
      \sum_m \, A^{m k_2}_{n_2 l_2 n_4 l_4}
         \left [ \sqrt{\frac{\pi}{2}} \widetilde{\mathcal{HO}}_{m k_2}(q')
   \right ]
   ,
\label{eq:intr2_2b}
\end{align}
where the coefficients $A^{m k_2}_{n_2 l_2 \, n_4 l_4}$ are given as
\begin{eqnarray}
   A^{m k_2}_{n_2 l_2 \, n_4 l_4} \, = \,
   \int_0^\infty & [{x'}_2^2 \, \mathrm{d}x'_2] \,
                      R_{n_2 l_2}({\scriptstyle \frac{1}{\sqrt{2}}} {x'}_2)
\label{3_12}
   \Bigl [
                   \mathcal{O}^{(2)} R_{n_4 l_4}({\scriptstyle \frac{1}{\sqrt{2}}} {x'}_2)
                   \Bigr ]
                                      \mathcal{HO}_{m k_2}({x'}_2)
   \> .
\end{eqnarray}
Using Equations (\ref{eq:intr1_2b}) and (\ref{eq:intr2_2b}), the
radial part of the two-body matrix elements becomes
\begin{eqnarray}
   &
   R^{\alpha \, \kappa; \, k_1 k_2}_{n_1 l_1 \, n_2 l_2 \, ; \, n_3 l_3 \, n_4 l_4}
   \! = & \frac{1}{(b \sqrt{2})^3}
              \sum_n \, A^{n k_1}_{n_1 l_1 \, n_3 l_3}
                         \sum_m \, A^{m k_2}_{n_2 l_2 \, n_4 l_4}
\label{3_13}
   \nonumber \\ & \quad & \times
   \int_0^\infty \, [{q'}^2 \, d{q'}] \,
                                    v_{\alpha \, \kappa}({\scriptstyle \frac{\sqrt{2}}{b}} \, q') \,
                                                               \widetilde{\mathcal{ HO}}_{n k_1}(q') \,
                                                               \widetilde{\mathcal{ HO}}_{m k_2}(q')
   \> .
\end{eqnarray}


\subsection{Central interaction}

The central interaction depends only on the magnitude of the
relative distance~$r$ between the two particles, i.e.
\begin{eqnarray}
   V_C \, = & \ V_{c}(r) \, \ck{0}{r}
   \>.
\end{eqnarray}
The strategy to calculating the \emph{ph} matrix element
\[ \jjmeph{V_C} \>, \]
is emblematic for the rest of this section: (For simplicity, we will not concerns ourselves here with the radial degrees of freedom. The radial part of the matrix elements is calculated using the approach discussed in the previous section.) First, we use lemma~2 and separate $V_C$ into tensor-operator components which
depend on either $\vec r_1$ or $\vec r_2$. We have
\begin{eqnarray}
   V_C \, = & \,    \sum_{k} \, (2k+1) \,
      u^{(k k ; 0 0)} (r_1,r_2) \,    \left ( \ck{k}{r_1} \odot \ck{k}{r_2} \right )
   \>.
\end{eqnarray}
Second, we use lemma~1 to calculate the \emph{ph} matrix element
of the corresponding interaction. For the central interaction, this
procedure leads to
\begin{align}
   & \jjmeph{V_C} =
   \sgn{j_2 + j_4 +1} \
   u^{(\lambda \lambda ; 0 0)} (r_1,r_2) \,
   \nonumber \\ & \times \
      \spjme{1}{\ck{\lambda}{r_1}}{3} \,    \spjme{2}{\ck{\lambda}{r_2}}{4}
   \>,
\end{align}
which gives
\begin{align}
    &
    \jjmeph{V_C} =
    \sgn{j_2 + j_4 +1} \
    u^{(\lambda \lambda ; 0 0)} (r_1,r_2) \,
   \nonumber \\ & \times \
        \Ffac{1}{3}{\lambda} \, \ckrme{1}{\lambda}{3} \,
            \Ffac{2}{4}{\lambda} \, \ckrme{2}{\lambda}{4}
    \>.
\end{align}


\subsection{Spin-spin interaction}

By definition, the spin-spin interaction is introduced as
\begin{equation}
   V_S \, = \, V_s(r) \, \sigma_1 \cdot \sigma_2
   \>,
\end{equation}
where we can write
\begin{equation}
  \sigma_1\cdot\sigma_2 \, = \,   - \sqrt{3} \, \left [ \sigma_1 \otimes \sigma_2 \right ]^{(0)}
  \>.
\end{equation}
We perform the recoupling
\begin{align}
    &
    \Bigr [ \,
    \bigl [ \ck{k}{r_1} \otimes \ck{k}{r_2} \bigr ]^{(0)}
    \, \otimes \,
    \bigl [ \sigma_1 \otimes \sigma_2 \bigr ]^{(0)}
    \, \Bigr ]^{(0)}
\label{recoup_s}
    \\ \nonumber & =
    \frac{1}{\hat k \sqrt 3}
    \sum_{l} \, \sgn{l} \,     \Bigl ( \,
    \bigl [ \ck{k}{r_1} \otimes \sigma_1 \bigr ]^{(l)}
    \, \odot \,
    \bigl [ \ck{k}{r_2} \otimes \sigma_2 \bigr ]^{(l)}
    \, \Bigr )
    \>,
\end{align}
and obtain
\begin{align}
    & \jjmeph{V_S}
    \\ \nonumber
    & =
    \frac{\sgn{\lambda + j_2 + j_4}}{2\lambda+1} \
    \sum_k \, \sgn{k} \, (2k+1) \,
        u^{(k k ; 0 0)} (r_1,r_2) \,
    \\ \nonumber & \quad \quad \times \
        \Gfac{1}{3}{k}{\lambda} \, \ckrme{1}{k}{3} \,
        \Gfac{2}{4}{k}{\lambda} \, \ckrme{2}{k}{4}
    \>,
\end{align}
with $|\lambda-1| \leq k \leq \lambda+1$ .


\subsection{Tensor interaction}

The tensor interaction is defined as:
\begin{equation}
   V_{T} = V_{t}(r) \, S_{12}
   \>,
\end{equation}
where the operator $S_{12}$ defined in Eq.~(\ref{S12_def}) is written now as
\begin{eqnarray}
  S_{12}
  \, = \,
  \sqrt{30} \,   \left [ \, \ck{2}{r} \otimes
  \left [ \sigma_1 \otimes \sigma_2 \right ]^{(2)} \, \right ]^{(0)}
  \>.
\end{eqnarray}
Using a recoupling scheme similar to Eq.~(\ref{recoup_s}),
\begin{align}
    &
    \Bigr [ \,
    \bigl [ \ck{k_1}{r_1} \otimes \ck{k_2}{r_2} \bigr ]^{(2)}
    \, \otimes \,
    \bigl [ \sigma_1 \otimes \sigma_2 \bigr ]^{(2)}
    \, \Bigr ]^{(0)}
\label{recoup_te}
    \\ \nonumber & = \
    - \ (-)^{k_2} \, \sqrt{5} \
    \sum_{k} \
    \wsj{k_1}{k_2}{2}{1}{1}{k} \,
    \Bigl ( \,
    \bigl [ \ck{k_1}{r_1} \otimes \sigma_1 \bigr ]^{(k)}
    \, \odot \,
    \bigl [ \ck{k_2}{r_2} \otimes \sigma_2 \bigr ]^{(k)}
    \, \Bigr )
    \>,
\end{align}
we obtain
\begin{align}
    & \jjmeph{V_T} =
    \frac{\sgn{j_2 + j_4 +1}}{2\lambda+1}
    \\ \nonumber
    & \times
    \sqrt{6} \,
    \sum_{k_1} \ (2k_1+1) \,
        \Gfac{1}{3}{k_1}{\lambda} \, \ckrme{1}{k_1}{3} \,
    \sum_{k_2} \, \mathrm{i}^{k_1+k_2} \, (2k_2+1) \,
            \\ \nonumber &
    \times
        u^{(k_1 k_2 ; 2 0)} (r_1,r_2)
        \cg{k_1}{0}{k_2}{0}{2}{0}
            \wsj{k_1}{k_2}{2}{1}{1}{\lambda}
                \Gfac{2}{4}{k_2}{\lambda} \ckrme{2}{k_2}{4}
    \>.
\end{align}


\subsection{Spin-orbit interaction}
\label{ls}

The spin-orbit interaction is given by:
\begin{equation}
  V_{LS} \ = \ V_{ls}(r) \, \mathbf{\ell} \cdot \mathbf{S}
  \>,
\end{equation}
where the orbital angular momentum operator and the total spin
operator are defined as
\begin{eqnarray}
   \vec{\ell}
   \ = \
   \frac{1}{2} \ \vec{r} \times (\vec{p_1} - \vec{p_2})
   \>,
   \qquad  \mathrm{and} \qquad
   \vec{S}
   \ = \
   \frac{1}{2} \ (\vec{\sigma_1} + \vec{\sigma_2})
   \>,
\end{eqnarray}
respectively. Using the tensor operator properties:
\begin{eqnarray}
   \vec r_m \ = \ r \ \ckq{1}{m}{r} \>,
   \qquad  \mathrm{and} \qquad
   (\vec{x} \times \vec{y} )_m
   \ = \
   - \mathrm{i} \ \sqrt{2} \
   \displaystyle \left [ x \otimes y \right ]^{(1)}_m
   \>,
\end{eqnarray}
we can write
\begin{eqnarray}
    \mathbf{\ell} & =
    \frac{\hbar}{\sqrt{2}}  \ r
    \left[ \ \ck{1}{r} \otimes (\grd_2 - \grd_1) \ \right]^{(1)}
    \> .
\end{eqnarray}
Then, we use lemma~2, and obtain:
\begin{align}
   &
   r V_{ls}(r) \, \ck{1}{r}
   \ = \
   \sum_{k_{1} k_{2}} \ \mathrm{i}^{k_2 - k_1 - 1} \
   \\ \nonumber &
   \times \,
   \frac{(2 k_1 + 1)(2 k_2 + 1)}{3 \sqrt{2}} \, \cg{k_1}{0}{k_2}{0}{1}{0} \,
   u^{(k_1 k_2, 1 1)} (r_1,r_2)
   \Bigl [ \cka{k_1} \otimes \ckb{k_2} \Bigr ]^{(1)}
   \>.
\end{align}
Recoupling the  $\cka{k_1}$ and $\ckb{k_2}$ operators with the
appropriate gradient operators, we obtain:
\begin{align}
    &
    V_{ls}(r) \, \mathbf{\ell}
    =
    \sum_{k_{1} k_{2}} \
         \mathrm{i}^{k_2 - k_1 - 1} \
         (2 k_1 + 1)\, (2 k_2 + 1) \
         \cg{k_1}{0}{k_2}{0}{1}{0} \
\label{eq:l_1}
    \\ \nonumber &
    \times \
    \sum_{k} \
         \frac{\hat k}{\sqrt 6} \
         \wsj{k_1}{k_2}{1}{1}{1}{k}
         u^{(k_1 k_2, 1 1)} (r_1,r_2) \
         \biggl \{
         - \Bigl [ \cka{k_1} \otimes
         \bigl [ \ckb{k_2} \otimes \grd_2 \bigr ]^{(k)} \Bigr ]^{(1)}
    \\ \nonumber & \qquad \qquad \qquad \qquad
         + \sgn{k+k_2+1}
         \Bigl [ \bigl [ \cka{k_1} \otimes \grd_1 \bigr ]^{(k)}
         \otimes \ckb{k_2} \Bigr ]^{(1)}
         \biggr \}
         \>.
\end{align}
Next, using the triangle conditions for the angular momentum
arguments of the Wigner $6j$ symbol, we derive the conditions: $|
k_1 - k_2 | \leq 1 \leq k_1 + k_2$, $| k_2 - 1 | \leq k \leq k_2 + 1$
and $| k_1 - 1 | \leq k \leq k_1 + 1$. These conditions together with the condition that $k_1+k_2$ is odd, derived from the Clebsch-Gordan coefficient in~(\ref{eq:l_1}),  can be satisfied only
if $k$ equals either $k_1$ or $k_2$. Therefore, we can write
Eq.~(\ref{eq:l_1}) as
\begin{align}
     &
     V_{ls}(r) \, \mathbf{\ell}
     =
     \frac{1}{\sqrt{2}}\
         \sum_{k_{1} k_{2}} \ \mathrm{i}^{k_1 + k_2 + 1} \
         (2 k_1 + 1)(2 k_2 + 1) \
         \cg{k_2}{0}{1}{0}{k_1}{0} \wsj{k_1}{k_2}{1}{1}{1}{k_1}
    \nonumber \\ & \qquad
    \times
    \biggl \{
    - \
         u^{(k_1 k_2, 1 1)} (r_1,r_2)
         \Bigl [ \cka{k_1} \otimes
         \bigl [ \ckb{k_2} \otimes \grd_2 \bigr ]^{(k_1)} \Bigr ]^{(1)}
    \nonumber \\ & \qquad \qquad
    \ + \
         u^{(k_1 k_2, 1 1)} (r_1,r_2)
         \Bigl [ \bigl [ \cka{k_1} \otimes \grd_1 \bigr ]^{(k_1)}
         \otimes \ckb{k_2} \Bigr ]^{(1)}
    \nonumber \\ & \qquad \qquad
    \ + \
         u^{(k_2 k_1, 1 1)} (r_1,r_2)
         \Bigl [ \cka{k_2} \otimes
         \bigl [ \ckb{k_1} \otimes \grd_2 \bigr ]^{(k_1)} \Bigr ]^{(1)}
    \nonumber \\ & \qquad \qquad
\label{eq:l_2}
    \ + \
         u^{(k_2 k_1, 1 1)} (r_1,r_2)
         \Bigl [ \bigl [ \cka{k_2} \otimes \grd_1 \bigr ]^{(k_1)}
         \otimes \ckb{k_1} \Bigr ]^{(1)}
    \biggr \}
    \> ,
\end{align}
with $k_1+k_2+1$ even.

Taking the inner products of Eq.~(\ref{eq:l_2}) with $\mathbf{S}$,
and using the following identities valid for operators $A^{(k)}$ and $B^{(\ell)}$ that commute with all components~$\sigma_i$
\begin{eqnarray}
   \biggl (
   \Bigl [ A^{(k)}_{(1)} \, \otimes \, B^{(\ell)}_{(2)} \Bigr ]^{(1)}
   \ \odot \ \sigma_1
   \biggr )
   & = &
   \sgn{k} \ \frac{\sqrt 3}{\hat \ell} \
   \biggl (
   \Bigl [ A^{(k)}_{(1)} \, \otimes \, \sigma_1 \Bigr ]^{(\ell)}
   \, \odot \, B^{(\ell)}_{(2)}
   \biggr )
   \>,
\label{eq:ls_id1}
   \\    \biggl (
   \Bigl [ A^{(k)}_{(1)} \, \otimes \, B^{(\ell)}_{(2)} \Bigr ]^{(1)}
   \ \odot \ \sigma_2
   \biggl )
   & = &
   \sgn{k+1} \ \frac{\sqrt 3}{\hat k} \
   \biggl (
   A^{(k)}_{(1)} \, \odot \,
   \Bigr [ B^{(\ell)}_{(2)} \, \otimes \, \sigma_2 \Bigr ]^{(k)}
   \biggr )
   \>,
\label{eq:ls_id2}
\end{eqnarray}
the operators in Eq.~(\ref{eq:l_2}) give the following contributions
to the \emph{ph}-coupled matrix element $\jjmeph{V_{LS}}$: \\
\noindent
\begin{align}
   &
   \biggl (
   \Bigl [ \cka{k_1}
           \otimes
           \bigl [ \ckb{k_2} \otimes \grd_2 \bigr ]^{(k_1)}
   \Bigr ]^{(1)}
   \, \odot \, S \biggl )
   \\ \nonumber &
   \Rightarrow \
   \delta_{k_1 \lambda} \
   \frac{\sgn{\lambda + j_2 + j_4 +1}}{\hat \lambda \, (2\lambda+1)} \
   \frac{\sqrt 3}{2}{} \
   \\ \nonumber & \quad \times
   \Bigl [
      \Gfac{1}{3}{\lambda}{\lambda} \ \ckrme{1}{\lambda}{3} \
      \Ffac{2}{4}{\lambda} \ CG_2(\lambda k_2, k_2 \lambda)
   \\ \nonumber & \qquad \
   -
      \Ffac{1}{3}{\lambda} \ \ckrme{1}{\lambda}{3} \
      \Gfac{2}{4}{\lambda}{\lambda} \ CG_2(\lambda k_2, k_2 \lambda)
   \Bigr ]
   \>,
\end{align}
and
\noindent
\begin{align}
   &
   \biggl (
   \Bigl [
         \bigl [ \cka{k_1} \otimes \grd_1 \bigr ]^{(k_1)}
         \otimes
         \ckb{k_2}
   \Bigr ]^{(1)}
   \, \odot \, S \biggr )
   \\ \nonumber &
   \Rightarrow \
   \frac{\sgn{k_1 + j_2 + j_4 +1}}{\hat \lambda \, (2\lambda+1)} \
   \frac{\sqrt 3}{2} \
   \\ \nonumber & \quad \times
   \Bigl [
      \delta_{k_2 \lambda} \
      \Gfac{1}{3}{k_1}{\lambda} \ CG_1(k_1 \lambda, k_1 k_1) \
      \Ffac{2}{4}{\lambda} \ \ckrme{2}{\lambda}{4}
   \\ \nonumber & \qquad \
   -
      \delta_{k_1 \lambda} \
      \Ffac{1}{3}{\lambda} \ CG_1(\lambda k_2, \lambda \lambda) \
      \Gfac{2}{4}{k_2}{\lambda} \ \ckrme{2}{k_2}{4}
   \Bigr ]
   \>,
\end{align}
and
\noindent
\begin{align}
   &
   \biggl (
   \Bigl [ \cka{k_2} \otimes
         \bigl [ \ckb{k_1} \otimes \grd_2 \bigr ]^{(k_1)} \Bigr ]^{(1)}
   \, \odot \, S \biggr )
   \\ \nonumber &
   \Rightarrow \
   \frac{\sgn{k_2 + j_2 + j_4 +1}}{\hat \lambda \, (2\lambda+1)} \
   \frac{\sqrt 3}{2} \
   \\ \nonumber & \quad \times
   \Bigl [
      \delta_{k_1 \lambda} \
      \Gfac{1}{3}{k_2}{\lambda} \ \ckrme{1}{k_2}{3} \
      \Ffac{2}{4}{\lambda} \ CG_2(k_2 \lambda, \lambda \lambda)
   \\ \nonumber & \qquad \
   -
      \delta_{k_2 \lambda} \
      \Ffac{1}{3}{\lambda} \ \ckrme{1}{\lambda}{3} \
      \Gfac{2}{4}{k_1}{\lambda} \ CG_2(\lambda k_1, k_1 k_1)
   \Bigr ]
   \>,
\end{align}
and
\noindent
\begin{align}
   &
   \biggl (
   \Bigl [
         \bigl [ \cka{k_2} \otimes \grd_1 \bigr ]^{(k_1)}
         \otimes \ckb{k_1}
   \Bigr ]^{(1)}
   \, \odot \, S \biggr )
   \\ \nonumber &
   \Rightarrow \
   \delta_{k_1 \lambda} \
   \frac{\sgn{\lambda + j_2 + j_4 +1}}{\hat \lambda \, (2\lambda+1)} \
   \frac{\sqrt 3}{2}{} \
   \\ \nonumber & \quad \times
   \Bigl [
      \Gfac{1}{3}{\lambda}{\lambda} \ CG_1(k_2 \lambda, k_2 \lambda) \
      \Ffac{2}{4}{\lambda} \ \ckrme{2}{\lambda}{4}
   \\ \nonumber & \qquad \
   -
      \Ffac{1}{3}{\lambda} \ CG_1(k_2 \lambda, k_2 \lambda) \
      \Gfac{2}{4}{\lambda}{\lambda} \ \ckrme{2}{\lambda}{4}
   \Bigr ]
   \>,
\end{align}
where we have introduced the notations
\begin{align}
   &
   CG_1(k_1 k_2, l_1 l_2) =
   \cga{k_1}{k_2}{1}{1}{l_1}{l_2}
   \>,
\label{cg1}
   \\ &
   CG_2(k_1 k_2, l_1 l_2) =
   \cgb{k_1}{k_2}{1}{1}{l_1}{l_2}
\label{cg2}
   \>.
\end{align}


\subsection{$\mathbf{\ell}^2$ interaction}
\label{l2}

The $\mathbf{\ell}^2$ interaction is defined as
\begin{equation}
   V_{L2} \ = \ V_{l2}(r) \, \mathbf{\ell}^2
   \>,
\end{equation}
where we can write
\begin{eqnarray}
    \mathbf{\ell}^2
    = & - \sqrt{3} \ [ \, \mathbf{\ell} \otimes \mathbf{\ell} \, ]^{(0)}
    \>.
\end{eqnarray}
It is useful to discuss this as a particular case of the more
general operator
\begin{eqnarray}
    \bigl [ \mathbf{\ell} \otimes \mathbf{\ell} \bigr ]^{(j)}
    \ = \
    - \ \frac{1}{2} \
           \left [ \, [ \, r \otimes ( p_1 - p_2 ) \, ]^{(1)}
           \otimes
           [ \, r \otimes ( p_1 - p_2 ) \, ]^{(1)} \, \right ]^{(j)}
    \>.
\label{rr}
\end{eqnarray}
We begin by using the definition of the spherical components of an
arbitrary vector, $a$, i.e. (\cite{Edmonds}, 5.9.4)
\begin{equation}
   a_{\pm 1} = \mp \frac{1}{\sqrt{2}} \ (a_x \pm \mathrm{i} \, a_y) \, ; \ \ \ a_0 = a_z
   \>.
\end{equation}
Then, we can show that
\begin{equation}
    \comm{r_{m}}{p_{n}} \ = \ \mathrm{i} \hbar \ \sgn{m} \ \del{n,}{-m}
    \>,
\end{equation}
and
\begin{eqnarray}
    \comm{(r_1 - r_2)_{m}}{(p_1 - p_2)_{n}} & = &
    2 \, \mathrm{i} \hbar \ \sgn{m} \ \del{n,}{-m}
    \> .
\end{eqnarray}
We can change the coupling scheme and combine the two $\vec{r}\, $s,
in Eq.~(\ref{rr}), into a single tensor operator dependent on the
relative-coordinate unit vector:
\begin{align}
   &
   \Bigl [ \, \bigl [ \, r \otimes ( p_1 - p_2 ) \bigr ]^{(1)}
   \otimes
   \bigl [ \, r \otimes ( p_1 - p_2 ) \bigr ]^{(1)} \, \Bigr ]_m^{(j)}
\label{rprp}
   \\ \nonumber & =
   3 \,  r^2 \,
   \sum_{\kappa \kappa'}
       \hat \kappa  \hat \kappa'
       \wnj{1}{1}{1}{1}{1}{1}{\kappa}{\kappa'}{j}
       \cg{1}{0}{1}{0}{\kappa}{0}
   \left [  \ck{\kappa}{r}
   \otimes
   [  ( p_1 - p_2 ) \otimes ( p_1 - p_2 ) ]^{(\kappa')}  \right ]_m^{(j)}
   \\ \nonumber & \qquad
   + \ 6 \, \mathrm{i} \ r\ \sgn{-j} \
   \wsj{1}{1}{1}{1}{1}{j}  \
   \left [ \ \ck{1}{r} \otimes ( p_1 - p_2 ) \ \right ]_m^{(j)}
   \>.
\end{align}
Because of the cross product property, $\vec{p}_m \times \vec{p}_n = 0$,
$\kappa$ and $\kappa'$ cannot be equal to 1. (The parameter $\kappa$ is also restricted to even values only because of the Clebsch-Gordan coefficient in~(\ref{rprp}).) Then, we can write
\begin{align}
    \bigl [ \mathbf{\ell} \otimes \mathbf{\ell} \bigr ]^{(j)}
\label{llj}
    = & \
    \frac{3}{2} \ r^2
    \!\!\! \sum_{\kappa \kappa' = 0,2} \!\!\!
        \hat \kappa \, \hat \kappa' \
        \wnj{1}{1}{1}{1}{1}{1}{\kappa}{\kappa'}{j} \
        \cg{1}{0}{1}{0}{\kappa}{0} \
    \\ \nonumber & \qquad \qquad \times \
    \Bigl [ \, \ck{\kappa}{r}
    \otimes
    \bigl [ \, ( \grd_1 - \grd_2 )
    \otimes ( \grd_1 - \grd_2 \, ) \bigr ]^{(\kappa')} \, \Bigr ]^{(j)}
    \\ \nonumber &
    - \ 3 \ r \ \sgn{j} \
    \wsj{1}{1}{1}{1}{1}{j} \
    \Bigl [ \ck{1}{r} \otimes ( \grd_1 - \grd_2 \, ) \Bigr ]^{(j)}
    \> .
\end{align}
In the particular case of the $\mathbf{\ell}^2$ interaction, the
rank $j$ is equal to 0, and we can use the symmetry properties of the Wigner
$6j$ and $9j$ symbols (\cite{Edmonds}, 6.4.14 and 6.3.2) to obtain the tensor-product form of $\mathbf{\ell}^2$ as
\begin{align}
    \mathbf{\ell}^2 = & \,
    \frac{3}{2} \, r^2
    \!\! \sum_{\kappa = 0,2} \!\!
        \hat \kappa
        \wsj{1}{1}{1}{1}{1}{\kappa}
        \cg{1}{0}{1}{0}{\kappa}{0}
    \Bigl [ \ck{\kappa}{r}
    \! \otimes \!
    \bigl [ ( \grd_1 - \grd_2 )
    \! \otimes \! ( \grd_1 - \grd_2 ) \bigl ]^{(\kappa)} \Bigr ]^{(0)}
    \nonumber \\ &
    - \ \sqrt{3} \ r \
    \Bigl [ \ck{1}{r} \otimes \bigl ( \grd_1 - \grd_2 \, \bigr ) \Bigr ]^{(0)}
    \> .
\label{eq:l2_1}
\end{align}
We expand the unnormalized spherical harmonics $\ck{\kappa}{r}$ and
$\ck{1}{r}$ using lemma~2, to obtain
\begin{align}
    &
    V_{l2}(r) \, \mathbf{\ell}^2
    =
    \frac{3}{2}
    \sum_{\kappa = 0,2} \frac{1}{\hat \kappa}
    \wsj{1}{1}{1}{1}{\kappa}{1}
\label{eq:l2_2}
    \\ \nonumber & \ \times
    \cg{1}{0}{1}{0}{\kappa}{0}
    \sum_{k_{1} k_{2}} \mathrm{i}^{k_2 - k_1 - \kappa}
    (2 k_1 + 1)(2 k_2 + 1) \cg{k_1}{0}{k_2}{0}{\kappa}{0}
    \\ \nonumber & \ \times
    u^{(k_1 k_2, \kappa 2)} (r_1,r_2)
    \Bigl [ \, \bigl [ \cka{k_1} \otimes \ckb{k_2} \bigr ]^{(\kappa)}
    \! \otimes \!
    \bigl [ ( \grd_1 - \grd_2 )
    \! \otimes \! ( \grd_1 - \grd_2 ) \bigr ]^{(\kappa)} \Bigr ]^{(0)}
    \\ \nonumber &
    - \ \frac{1}{\sqrt{3}} \
    \sum_{k_{1} k_{2}} \, \mathrm{i}^{k_2 - k_1 - 1} \,
    (2 k_1 + 1)(2 k_2 + 1) \, \cg{k_1}{0}{k_2}{0}{1}{0} \
    u^{(k_1 k_2, 1 1)} (r_1,r_2)
    \\ \nonumber & \ \times
    \biggl \{
    \Bigl [ \bigl [ \cka{k_1} \! \otimes \! \ckb{k_2} \bigr ]^{(1)}
    \otimes \grd_{2} \Bigr ]^{(0)}
    \!\! -
    \Bigl [ \bigl [ \cka{k_1} \! \otimes \! \ckb{k_2} \bigr ]^{(1)}
    \! \otimes \! \grd_{1} \Bigr ]^{(0)}
    \biggr \}
    \> .
\end{align}
The operators in Eq.~(\ref{eq:l2_2}) give rise to contributions to the matrix element
$\jjmeph{V_{L2}}$
similar to the
contributions of the $\mathbf{\ell}^2 \, \bigl ( \sigma_1 \cdot
\sigma_2 \bigr )$ and quadrupole spin-orbit operators discussed in
the subsequent sections. Therefore, it is useful to base the
calculation of the $\mathbf{\ell}^2$ matrix element on several
general results.

First, let us discuss the case of the matrix elements involving only
one gradient operator in Eq.~(\ref{eq:l2_2}), which  are very
similar to those discussed in the case of the spin-orbit
interaction. We have the general identities:
\begin{align}
   &
   \Bigl [
       \bigl [ \cka{k_1} \otimes \ckb{k_2} \bigr ]^{(1)}
       \otimes \grd_2 \Bigr ]^{(j)}
\label{eq:l2_Dj}
   \\ \nonumber & =
   \sgn{j} \, \sqrt{3} \
   \sum_k \ \hat k \
            \wsj{k_1}{k_2}{1}{1}{j}{k}
   \Bigl [ \cka{k_1} \otimes
         \bigl [ \ckb{k_2} \otimes \grd_2 \bigr ]^{(k)} \Bigr ]^{(j)}
   \>,
\end{align}
and
\begin{align}
   &
   \Bigl [
       \bigl [ \cka{k_1} \otimes \ckb{k_2} \bigr ]^{(1)}
       \otimes \grd_1 \Bigr ]^{(j)}
\label{eq:l2_Ej}
   \\ \nonumber & =
   \sgn{k_2} \, \sqrt{3} \
   \sum_k \ \sgn{k} \, \hat k \
            \wsj{k_2}{k_1}{1}{1}{j}{k}
   \Bigl [ \bigl [ \cka{k_1} \otimes \grd_1 \bigr ]^{(k)}
           \otimes
           \ckb{k_2} \Bigr ]^{(j)}
   \>.
\end{align}
For the $\mathbf{\ell}^2$ interaction we are only interested in $j=0$. We obtain:\\
\noindent
\begin{align}
   &
   \Bigl [
       \bigl [ \cka{k_1} \otimes \ckb{k_2} \Bigr ]^{(1)}
       \otimes \grd_2 \Bigr ]^{(0)}
\label{eq:l2_D}
   \\ \nonumber &
   \Rightarrow
   \delta_{k_1 \lambda}
   \frac{\sgn{\lambda + j_2 + j_4 +1}}{\hat \lambda \, (2\lambda+1)}
   \Ffac{1}{3}{\lambda} \ckrme{1}{k_1}{3}
   \Ffac{2}{4}{\lambda} CG_2(\lambda k_2, k_2 \lambda)
   \>,
\end{align}
and
\noindent
\begin{align}
   &
   \Bigl [
       \bigl [ \cka{k_1} \otimes \ckb{k_2} \bigr ]^{(1)}
       \otimes \grd_1 \Bigr ]^{(0)}
\label{eq:l2_E}
   \\ \nonumber &
   \Rightarrow
   \delta_{k_2 \lambda}
   \frac{\sgn{\lambda + j_2 + j_4 +1}}{\hat \lambda \, (2\lambda+1)}
   \Ffac{1}{3}{\lambda} CG_1(k_1 \lambda, k_1 \lambda)
   \Ffac{2}{4}{\lambda} \ckrme{2}{\lambda}{4}
   \>.
\end{align}
Next, we multiply the operator expressions in the first sum in
Eq.~(\ref{eq:l2_2}) to obtain
\begin{align}
    \Bigl [ \, \bigl [ \cka{k_1} \otimes & \ckb{k_2} \bigr ]^{(\kappa)}
    \otimes
    \bigl [ \, ( \grd_1 - \grd_2 )
    \otimes ( \grd_1 - \grd_2 \, ) \bigr ]^{(\kappa)} \, \Bigr ]^{(0)}
\label{eq:r2_terms}
    \\ \nonumber = &
    \Bigl [ \bigl [ \cka{k_1} \otimes \ckb{k_2} \bigr ]^{(\kappa)}
    \otimes
    \bigl [ \grd_2 \otimes \grd_2 \bigr ]^{(\kappa)} \Bigr ]^{(0)}
    \\ \nonumber & \quad
    - 2
    \Bigl [ \bigl [ \cka{k_1} \otimes \ckb{k_2} \bigr ]^{(\kappa)}
    \otimes
    \bigl [ \grd_1 \otimes \grd_2 \bigr ]^{(\kappa)} \Bigr ]^{(0)}
    \\ \nonumber & \quad
    +
    \Bigl [ \bigl [ \cka{k_1} \otimes \ckb{k_2} \bigr ]^{(\kappa)}
    \otimes
    \bigl [ \grd_1 \otimes \grd_1 \bigr ]^{(\kappa)} \Bigr ]^{(0)}
    \>.
\end{align}
Now, we have to change the coupling and separate the operators
depending on the coordinates of the first particle, from the
operators depending on the coordinates of the second particle.
Hence, the matrix elements involving two gradient operators in
Eq.~(\ref{eq:l2_2}), are calculated using the following identities:
\begin{align}
    &
    \Bigl [ \,
    \bigl [ \cka{k_1} \otimes \ckb{k_2} \bigr ]^{(\kappa)}
    \otimes
    \bigl [ \, \grd_2 \otimes \grd_2 \bigr ]^{(\kappa')} \,
    \Bigr ]^{(j)}
\label{eq:l2_Aj}
    \\ \nonumber &
    =
    \sum_k
    \sgn{k_1 + k_2 + \kappa' + j}
    \hat \kappa  \hat k
    \wsj{k_1}{k_2}{\kappa}{\kappa'}{j}{k}
    \Bigl [
    \cka{k_1} \! \otimes \!
    \Bigl [ \ckb{k_2}
    \! \otimes \!
    \bigl [  \grd_2 \! \otimes \! \grd_2 \bigr ]^{(\kappa')} \Bigr ]^{(k)}
    \Bigr ]^{(j)}
    \>,
\end{align}
and
\begin{align}
    &
    \Bigl [ \,
    \bigl [ \cka{k_1} \otimes \ckb{k_2} \bigr ]^{(\kappa)}
    \otimes
    \bigl [ \, \grd_1 \otimes \grd_2 \bigr ]^{(\kappa')} \,
    \Bigr ]^{(j)}
\label{eq:l2_Bj}
    \\ \nonumber &
    =
    \sum_{k k'} \,
    \hat \kappa \, \hat \kappa' \, \hat k \, \hat k' \,
    \wnj{k_1}{k_2}{\kappa}{1}{1}{\kappa'}{k}{k'}{j}
    \Bigl [
    \bigl [ \cka{k_1} \otimes \grd_1 \bigr ]^{(k)} \,
    \otimes
    \bigl [ \ckb{k_2} \otimes \grd_2 \bigr ]^{(k')}
    \Bigr ]^{(j)}
    \>,
\end{align}
and
\begin{align}
    &
    \Bigl [ \,
    \bigl [ \cka{k_1} \otimes \ckb{k_2} \bigr ]^{(\kappa)}
    \otimes
    \bigl [ \, \grd_1 \otimes \grd_1 \bigr ]^{(\kappa')} \,
    \Bigr ]^{(j)}
\label{eq:l2_Cj}
    \\ \nonumber &
    =
    \sum_k
    \sgn{k_2 + \kappa' - \kappa + k}
    \hat \kappa  \hat k
    \wsj{k_2}{k_1}{\kappa}{\kappa'}{j}{k}
    \Bigl [  \Bigl [
    \cka{k_1} \! \otimes \!
    \bigl [  \grd_1 \! \otimes \! \grd_1 \bigr ]^{(\kappa')} \Bigr ]^{(k)}
    \otimes
    \ckb{k_2}
    \Bigr ]^{(j)}
    \>.
\end{align}
Again, for the $\mathbf{\ell}^2$ interaction, we are only interested
in the case
$j=0$. We obtain:\\
\noindent
\begin{align}
    &
    \Bigl [ \,
    \bigl [ \cka{k_1} \otimes \ckb{k_2} \bigr ]^{(\kappa)}
    \otimes
    \bigl [ \, \grd_2 \otimes \grd_2 \bigr ]^{(\kappa)} \,
    \Bigr ]^{(0)}
\label{eq:l2_A}
    \\ \nonumber &
    \Rightarrow
    \delta_{k_1 \lambda}
    \frac{\sgn{\lambda + j_2 + j_4 +1}}{\hat \lambda \, (2\lambda+1)}
    \Ffac{1}{3}{\lambda} \ckrme{1}{\lambda}{3}
    \\ \nonumber &
    \qquad \qquad \qquad \times \
    \Ffac{2}{4}{\lambda} CGG_2(\lambda k_2 \kappa, k_2 \kappa \lambda)
    \>,
\end{align}
and
\noindent
\begin{align}
    &
    \Bigl [ \,
    \bigl [ \cka{k_1} \otimes \ckb{k_2} \bigr ]^{(\kappa)}
    \otimes
    \bigl [ \, \grd_1 \otimes \grd_2 \bigr ]^{(\kappa)} \,
    \Bigr ]^{(0)}
\label{eq:l2_B}
    \\ \nonumber &
    \Rightarrow
    \frac{\sgn{k_2 + j_2 + j_4}}{2\lambda+1} \
    \hat \kappa \,
    \wsj{k_1}{k_2}{\kappa}{1}{1}{\lambda}
    \\ \nonumber &
    \qquad \qquad \qquad \times \
    \Ffac{1}{3}{\lambda} \Ffac{2}{4}{\lambda}
    CGCG(k_1 k_2 \kappa, k_1 \lambda k_2 \lambda)
    \>,
\end{align}
and
\noindent
\begin{align}
    &
    \Bigl [ \,
    \bigl [ \cka{k_1} \otimes \ckb{k_2} \bigl ]^{(\kappa)}
    \otimes
    \bigl [ \, \grd_1 \otimes \grd_1 \bigr ]^{(\kappa)} \,
    \Bigr ]^{(0)}
\label{eq:l2_C}
    \\ \nonumber &
    \Rightarrow
    \delta_{k_2 \lambda}
    \frac{\sgn{\lambda + j_2 + j_4 +1}}{\hat \lambda \, (2\lambda+1)}
    \Ffac{1}{3}{\lambda} CGG_1(k_1 \lambda \kappa, k_1 \kappa \lambda)
    \\ \nonumber &
    \qquad \qquad \qquad \times \
    \Ffac{2}{4}{\lambda} \ckrme{2}{\lambda}{4}
    \>.
\end{align}
Here we have introduced the notations
\begin{align}
   &
   CGG_1(k_1 k_2 \kappa, k l \lambda)
   =
   \cgga{k_1}{k_2}{2}{\kappa}{k}{l}{\lambda}
\end{align}
\begin{align}
   &
   CGG_2(k_1 k_2 \kappa, k l \lambda)
   =
   \cggb{k_1}{k_2}{2}{\kappa}{k}{l}{\lambda}
\end{align}
and
\begin{align}
   CGCG(k_1 k_2 \kappa, l_1 l_2 \lambda_1 \lambda_2)
   = & \
   \langle 
   l_1 \, \| \,
        u^{(k_1, 2 \kappa)} (r_1) \,
        \left[ C^{(l_1)} (\hat{r_1}) \otimes \grd_1
        \right ]^{(l_2)}
        \, \| \,
        l_3 \rangle
   \\ \nonumber & \times \
   \langle 
   l_2 \, \| \,
        u^{(k_2, 2 \kappa)} (r_2) \,
        \left[ C^{(\lambda_1)} (\hat{r_2}) \otimes \grd_2
        \right ]^{(\lambda_2)}
        \, \| \,
        l_4 \rangle
   \>.
\end{align}


\subsection{$\mathbf{\ell}^2 \, \bigl ( \sigma_1 \cdot \sigma_2 \bigr )$ interaction}
\label{l2ss}

The $\mathbf{\ell}^2 \, \bigl ( \sigma_1 \cdot \sigma_2 \bigr )$
interaction is given by:
\begin{eqnarray}
    V_{L2S} \ = \
    V_{l2s}(r) \, \mathbf{\ell}^2 \, \bigl ( \sigma_1 \cdot \sigma_2 \bigr ) & = &
    \left ( - \, \sqrt{3} \right ) \
    V_{l2s}(r) \, \mathbf{\ell}^2 \,
    \left [ \sigma_1 \otimes \sigma_2 \right]^{(0)}
    \>.
\end{eqnarray}
Since the $\mathbf{\ell}^2 \, \bigl ( \sigma_1 \cdot \sigma_2 \bigr
)$ interaction differs from the $\mathbf{\ell}^2$ interaction only
through the spin part, $\left ( \sigma_1\cdot\sigma_2 \right )$, we
can use Eq.~(\ref{eq:l2_2}) and add the corresponding spin
interaction. We have:
\begin{align}
    &
    V_{l2s}(r) \, \mathbf{\ell}^2 \, \left ( \sigma_1\cdot\sigma_2 \right )
    =
    \frac{3}{2} \
    \sum_{\kappa = 0,2} \frac{1}{\hat \kappa}
    \wsj{1}{1}{1}{1}{\kappa}{1}
\label{eq:l2ss_2}
    \\ \nonumber & \ \times
    \cg{1}{0}{1}{0}{\kappa}{0}
    \sum_{k_{1} k_{2}} \mathrm{i}^{k_2 - k_1 - \kappa} \,
    (2 k_1 + 1)(2 k_2 + 1) \, \cg{k_1}{0}{k_2}{0}{\kappa}{0}
    u^{(k_1 k_2, \kappa 2)} (r_1,r_2)
    \\ \nonumber & \ \times
    \Bigl [ \bigl [ \cka{k_1} \otimes \ckb{k_2} \bigr ]^{(\kappa)}
    \otimes
    \bigl [ ( \grd_1 - \grd_2 )
    \otimes ( \grd_1 - \grd_2 ) \bigr ]^{(\kappa)} \Bigr ]^{(0)}
    \left ( \sigma_1\cdot\sigma_2 \right )
    \\ \nonumber &
    - \ \frac{1}{\sqrt{3}} \
    \sum_{k_{1} k_{2}} \, \mathrm{i}^{k_2 - k_1 - 1} \,
    (2 k_1 + 1)(2 k_2 + 1) \, \cg{k_1}{0}{k_2}{0}{1}{0} \
    u^{(k_1 k_2, 1 1)} (r_1,r_2)
    \\ \nonumber & \ \times \,
    \biggl \{
    \Bigl [ \bigl [ \cka{k_1} \otimes \ckb{k_2} \bigr ]^{(1)}
    \otimes \grd_{2} \Bigr ]^{(0)}
    ( \sigma_1\cdot\sigma_2 )
    \\ \nonumber & \qquad \qquad \qquad \qquad
    -
    \Bigl [ \bigl [ \cka{k_1} \otimes \ckb{k_2} \bigr ]^{(1)}
    \otimes \grd_{1} \Bigr ]^{(0)}
    ( \sigma_1\cdot\sigma_2 )
    \biggr \}
    \> .
\end{align}
Accordingly, using Eqs.~(\ref{eq:l2_A}--\ref{eq:l2_C}) and
Eqs.~(\ref{eq:l2_D},\ref{eq:l2_E}) we obtain the
contributions to the \emph{ph} matrix element $\jjmeph{V_{L2S}}$:\\
\noindent
\begin{align}
    &
    \Bigl [ \,
    \bigl [ \cka{k_1} \otimes \ckb{k_2} \bigr ]^{(\kappa)}
    \otimes
    [ \grd_2 \otimes \grd_2 ]^{(\kappa)} \,
    \Bigr ]^{(0)} \
    \left ( \sigma_1\cdot\sigma_2 \right )
    \\ \nonumber &
    \Rightarrow \
   \frac{\sgn{\lambda + j_2 + j_4}}{\hat k_1 \, (2\lambda+1)} \
    \Gfac{1}{3}{k_1}{\lambda} \ \ckrme{1}{k_1}{3} \
    \\ \nonumber &
    \qquad \qquad \qquad \times \
    \Gfac{2}{4}{k_1}{\lambda} \ CGG_2(k_1 k_2 \kappa, k_2 \kappa k_1)
    \>,
\end{align}
and
\noindent
\begin{align}
    &
    \Bigl [ \,
    \bigl [ \cka{k_1} \otimes \ckb{k_2} \bigr ]^{(\kappa)}
    \otimes
    [ \, \grd_1 \otimes \grd_2 ]^{(\kappa)} \,
    \Bigr ]^{(0)} \
    \left ( \sigma_1\cdot\sigma_2 \right )
    \\ \nonumber &
    \Rightarrow \
   \frac{\sgn{\lambda + k_2 + j_2 + j_4 +1}}{2\lambda+1} \
    \hat \kappa \,
    \sum_{k} \ \sgn{k} \
    \wsj{k_1}{k_2}{\kappa}{1}{1}{k} \
    \\ \nonumber &
    \qquad \qquad \qquad \times \
    \Gfac{1}{3}{k}{\lambda} \ \Gfac{2}{4}{k}{\lambda} \
    CGCG(k_1 k_2 \kappa, k_1 k k_2 k)
    \>,
\end{align}
and
\noindent
\begin{align}
    &
    \Bigl [ \,
    \bigl [ \cka{k_1} \otimes \ckb{k_2} \bigr ]^{(\kappa)}
    \otimes
    [ \, \grd_1 \otimes \grd_1 ]^{(\kappa)} \,
    \Bigr ]^{(0)} \
    \left ( \sigma_1\cdot\sigma_2 \right )
    \\ \nonumber &
    \Rightarrow
   \frac{\sgn{\lambda + j_2 + j_4}}{\hat k_2 \, (2\lambda+1)} \
    \Gfac{1}{3}{k_2}{\lambda} \ CGG_1(k_1 k_2 \kappa, k_1 \kappa k_2) \
    \\ \nonumber &
    \qquad \qquad \qquad \times \
    \Gfac{2}{4}{k_2}{\lambda} \ \ckrme{2}{k_2}{4}
    \>,
\end{align}
and
\noindent
\begin{align}
   &
   \Bigl [
       \bigl [ \cka{k_1} \otimes \ckb{k_2} \bigr ]^{(1)}
       \otimes \grd_2 \Bigr ]^{(0)} \
   \left ( \sigma_1\cdot\sigma_2 \right )
   \\ \nonumber &
   \Rightarrow \
   \frac{\sgn{\lambda + j_2 + j_4}}{\hat k_1 \, (2\lambda+1)} \
   \Gfac{1}{3}{k_1}{\lambda} \ \ckrme{1}{k_1}{3} \
    \\ \nonumber &
    \qquad \qquad \qquad \times \
   \Gfac{2}{4}{k_1}{\lambda} \ CG_2(k_1 k_2, k_2 k_1)
   \>,
\end{align}
and
\noindent
\begin{align}
   &
   \Bigl [
       \bigl [ \cka{k_1} \otimes \ckb{k_2} \bigr ]^{(1)}
       \otimes \grd_1 \Bigr ]^{(0)} \
   \left ( \sigma_1\cdot\sigma_2 \right )
   \\ \nonumber &
   \Rightarrow \
   \frac{\sgn{\lambda + j_2 + j_4}}{\hat k_2 \, (2\lambda+1)} \
   \Gfac{1}{3}{k_2}{\lambda} \ CG_1(k_1 k_2, k_1 k_2) \
    \\ \nonumber &
    \qquad \qquad \qquad \times \
   \Gfac{2}{4}{k_2}{\lambda} \ \ckrme{2}{k_2}{4}
   \>.
\end{align}


\subsection{Quadrupole spin-orbit interaction}
\label{ls2}

The quadrupole spin-orbit interaction, $V_{LS2}$, is given as the
radial factor, $V_{ls2}(r)$, multiplying the operator $(\mathbf{\ell} \cdot \mathbf{S})^2$ defined in Eq.~(\ref{ls2_def}).
As discussed in Sec.~\ref{ops}, the $j=0,1$ components in Eq.~(\ref{ls2_def}) can be
incorporated by introducing modified radial amplitudes of the
spin-orbit, $\mathbf{\ell}^2$ and $\mathbf{\ell}^2 \, \bigl (
\sigma_1 \cdot \sigma_2 \bigr )$ interactions, see Eqs.~(\ref{l2_mod},\ref{l2s_mod},\ref{ls_mod}).
The only component of the interaction that we have not addressed yet
is the one in Eq.~(\ref{j2}), corresponding to $j=2$.

In general, we can use Eq.~(\ref{llj}) and write
\begin{align}
    &
    V_{ls2}(r) \, \left [ \mathbf{\ell} \otimes \mathbf{\ell} \right ]^{(j)}
    =
    \frac{3}{2} \, r^2
    \!\! \sum_{\kappa \kappa' = 0,2} \!
        \frac{\hat \kappa'}{\hat \kappa}
        \wnj{1}{1}{1}{1}{1}{1}{\kappa}{\kappa'}{j}
\label{eq:ls2_2}
    \\ \nonumber & \
    \times \,
        \cg{1}{0}{1}{0}{\kappa}{0}
    \sum_{k_{1} k_{2}} \mathrm{i}^{k_2 - k_1 - \kappa}
    (2 k_1 + 1)(2 k_2 + 1) \cg{k_1}{0}{k_2}{0}{\kappa}{0}
    u^{(k_1 k_2, 2 \kappa)} (r_1,r_2)
    \\ \nonumber & \
    \times \,
    \Bigl [
          \bigl [ \cka{k_1} \otimes \ckb{k_2} \bigr ]^{(\kappa)}
          \otimes
          \bigl [ ( \grd_1 - \grd_2 )
               \otimes ( \grd_1 - \grd_2 ) \bigr ]^{(\kappa')}
    \Bigr ]^{(j)}
    \\ \nonumber &
    - \ r \ \sgn{j} \
    \wsj{1}{1}{1}{1}{1}{j} \
    \sum_{k_{1} k_{2}} \, \mathrm{i}^{k_2 - k_1 - 1} \,
    (2 k_1 + 1)(2 k_2 + 1) \, \cg{k_1}{0}{k_2}{0}{1}{0} \
    \\ \nonumber & \qquad \qquad \qquad
    \times \
    u^{(k_1 k_2, 1 1)} (r_1,r_2)
    \Bigl [ \bigl [ \cka{k_1} \otimes \ckb{k_2} \bigr ]^{(1)}
          \otimes ( \grd_1 - \grd_2 \, )
    \Bigr ]^{(j)}
    \> .
\end{align}
Using Eqs.~(\ref{eq:l2_Aj}--\ref{eq:l2_Cj}) and
Eqs.~(\ref{eq:l2_Dj},\ref{eq:l2_Ej}), with $j=2$, we couple
$\cka{k}$ with $\grd_1$, and $\ckb{l}$ with $\grd_2$ dependent
operators, respectively.
The contributions to the
\emph{ph}-coupled matrix element $\jjmeph{V_{LS2}}$ are:
\noindent
\begin{align}
   &
   \biggl [
      \Bigl [
         \cka{k_1} \otimes
         \Bigl [
            \ckb{k_2}
            \otimes
            \bigl [ \grd_2 \otimes \grd_2 \bigr ]^{(\kappa')}
         \Bigr ]^{(k)}
      \Bigr ]^{(2)}
      \otimes
      \left [ \sigma_1 \otimes \sigma_2 \right ]^{(2)} \
   \biggl ]^{(0)}
   \\ \nonumber &
   \Rightarrow \
   \frac{\sgn{k + j_2 + j_4}}{2\lambda+1} \
   \sqrt{5} \
   \wsj{k_1}{k}{2}{1}{1}{\lambda} \
   \\ \nonumber & \ \times \,
   \Gfac{1}{3}{k_1}{\lambda} \, \ckrme{1}{k_1}{3} \,
   \Gfac{2}{4}{k}{\lambda} \, CGG_2(k_1 k_2 \kappa, k_2 \kappa' k)
   \>,
\end{align}
and
\noindent
\begin{align}
   &
   \biggl [
      \Bigl [
            \bigl [ \cka{k_1} \otimes \grd_1 \bigr ]^{(k)}
            \otimes
            \bigl [ \ckb{k_2} \otimes \grd_2 \bigr ]^{(k')}
      \Bigr ]^{(2)}
      \otimes
      \left [ \sigma_1 \otimes \sigma_2 \right ]^{(2)} \
   \biggl ]^{(0)}
   \\ \nonumber &
   \Rightarrow \
   \frac{\sgn{k' + j_2 + j_4 +1}}{2\lambda+1} \
   \sqrt{5} \
   \wsj{k}{k'}{2}{1}{1}{\lambda} \
   \\ \nonumber & \qquad \times \,
   \Gfac{1}{3}{k}{\lambda} \, \Gfac{2}{4}{k'}{\lambda} \,
   CGCG(k_1 k_2 \kappa, k_1 k k_2 k')
   \>,
\end{align}
and
\noindent
\begin{align}
   &
   \biggl [
      \Bigr [
         \Bigr [
            \cka{k_1}
            \otimes
            \bigl [ \grd_1 \otimes \grd_1 \bigr ]^{(\kappa')}
         \Bigr ]^{(k)}
         \otimes \ckb{k_2}
      \Bigr ]^{(2)}
      \otimes
      \left [ \sigma_1 \otimes \sigma_2 \right ]^{(2)} \
   \biggr ]^{(0)}
   \\ \nonumber &
   \Rightarrow \
   \frac{\sgn{k_2 + j_2 + j_4 +1}}{2\lambda+1} \
   \sqrt{5} \
   \wsj{k}{k_2}{2}{1}{1}{\lambda} \
   \\ \nonumber & \ \times \,
   \Gfac{1}{3}{k}{\lambda} \, CGG_1(k_1 k_2 \kappa, k_1 \kappa' k) \,
   \Gfac{2}{4}{k_2}{\lambda} \,  \ckrme{2}{k_2}{4}
   \>,
\end{align}
and
\noindent
\begin{align}
   &
   \biggl [
      \Bigl [
         \cka{k_1} \otimes
         \bigl [ \ckb{k_2} \otimes \grd_2 \bigr ]^{(k)}
      \Bigr ]^{(2)}
      \otimes
      \left [ \sigma_1 \otimes \sigma_2 \right ]^{(2)} \
   \biggr ]^{(0)}
   \\ \nonumber &
   \Rightarrow \
   \frac{\sgn{k + j_2 + j_4}}{2\lambda+1} \
   \sqrt{5} \
   \wsj{k_1}{k}{2}{1}{1}{\lambda} \
   \\ \nonumber & \qquad \times \,
   \Gfac{1}{3}{k_1}{\lambda} \, \ckrme{1}{k_1}{3} \,
   \Gfac{2}{4}{k}{\lambda} \, CG_2(k_1 k_2, k_2 k)
   \>,
\end{align}
and
\noindent
\begin{align}
   &
   \biggl [
      \Bigl [
         \bigl [ \cka{k_1} \otimes \grd_1 \bigr ]^{(k)}
         \otimes \ckb{k_2}
      \Bigr ]^{(2)}
      \otimes
      \bigl [ \sigma_1 \otimes \sigma_2 \bigr ]^{(2)} \
   \biggr ]^{(0)}
   \\ \nonumber &
   \Rightarrow \
   \frac{\sgn{k_2 + j_2 + j_4}}{2\lambda+1} \
   \sqrt{5} \
   \wsj{k}{k_2}{2}{1}{1}{\lambda} \
   \\ \nonumber & \qquad \times \,
   \Gfac{1}{3}{k}{\lambda} \, CG_1(k_1 k_2, k_1 k) \,
   \Gfac{2}{4}{k_2}{\lambda} \, \ckrme{2}{k_2}{4}
   \>.
\end{align}


\section{Isospin matrix-elements calculation}
\label{iso}

Independent of the functional form of the single-particle wave functions,
the isospin dependence of the matrix elements is worked out in
a proton-neutron basis representation. We find it convenient to introduce the
proton/neutron creation/destruction operators, $\tau_\pm$, in terms
of its Cartesian components, as
\begin{eqnarray}
   \tau_\pm = \frac{1}{2} \, (\tau_x \, \pm \, i \tau_y)
   \>, \quad
   \tau_0 = \tau_z
   \>,
\end{eqnarray}
such that
\begin{eqnarray*}
   \tau_+ \, | p \rangle & = \ 0
   \qquad \qquad
   \tau_+ \, | n \rangle & = \ | p \rangle
   \\
   \tau_-\, | n \rangle & = \ 0
   \qquad \qquad
   \tau_-\, | p \rangle & = \ | n \rangle
   \> .
\end{eqnarray*}
Reciprocally, we have
\begin{eqnarray}
   \tau_x & = & \tau_+ \, + \, \tau_-
   \>,
   \\
   \tau_y & = & \frac{1}{i} \ (\tau_+ \, - \, \tau_-)
   \>,
\end{eqnarray}
where the Cartesian components of the isospin operator are the usual
Pauli matrices
\begin{eqnarray}
   \tau_x \ = \
   \Bigl (
      \begin{array}{cc}
         0 & 1 \\          1 & 0
      \end{array}
   \Bigr )
   \>,
   \quad
   \tau_y \ = \
   \Bigl (
      \begin{array}{cc}
         0 & -i \\          i & 0
      \end{array}
   \Bigr )
   \>,
   \quad
   \tau_z \ = \
   \Bigl (
      \begin{array}{cc}
         1 & 0 \\          0 & -1
      \end{array}
   \Bigr )
   \>.
\end{eqnarray}
With these definitions, the expectation values of the various
isospin-dependent operators are computed,
in terms of the matrices
\begin{eqnarray}
   \langle \tau' \, | \, \tau_+ \, | \, \tau \rangle
   & =
   \left \lbrace
      \begin{array}{ll}
         1 & {\rm if} \ | \tau  \rangle = | n \rangle \ \mathrm{and}
                      \ | \tau' \rangle = | p \rangle
         \>,
         \\
         0 & {\rm otherwise}
         \>,
      \end{array}
   \right .
   \\
   \langle \tau' \, | \, \tau_- \, | \, \tau \rangle
   & =
   \left \lbrace
      \begin{array}{ll}
         1 & {\rm if} \ | \tau  \rangle = | p \rangle \ \mathrm{and}
                      \ | \tau' \rangle = | n \rangle
         \>,
         \\
         0 & {\rm otherwise}
         \>,
      \end{array}
   \right .
   \\
      \langle \tau' \, | \, \tau_0 \, | \, \tau \rangle
   & =
   \left \lbrace
      \begin{array}{ll}
         1 & {\rm if} \ | \tau  \rangle = | \tau' \rangle = | p \rangle
         \>,
         \\
         -1 & {\rm if} \ | \tau  \rangle = | \tau' \rangle = | n \rangle
         \>,
         \\
         0 & {\rm otherwise} \>.
      \end{array}
   \right .
\end{eqnarray}
In particular, we note the identity:
         \begin{equation}
            \tau_1 \cdot \tau_2
            \ = \
            2 \, ( \tau_{1 +} \tau_{2 -} \, + \, \tau_{1 -} \tau_{2 +} )
            \ + \
            \tau_{1 0} \tau_{2 0}
            \>.
         \end{equation}


\begin{table}[t]
\caption{\label{tab:dr}Convergence of matrix elements of the
 central interaction of the Argonne $v_{18}$
 potential calculated using the Moshinsky transformation-brackets
 approach outlined in Sec.~\ref{ops}.
 Here we study the evolution of the parameter $\varepsilon_{>x}^{\le y}$,
 which indicates how many matrix elements
 out of the total number of matrix elements, NME,
 that can be formed in a $^{16}$O-like model space of single-particle
 wave functions, $\mathcal{R}_{nl}$,
 with $l \le 6$ and $n \le 6$, are characterized by a relative error ranging between
 $x\%$ and $y\%$ when doubling the number of abscissas and weights in the
 Gauss-Hermite quadrature set, NGAUS,
 use to compute the radial integral~(\ref{radmat}).}
\centering
\begin{tabular}{rcrrrr}
\hline
NGAUS & NME & $\varepsilon_{> 1}$ &
                $\varepsilon_{> 0.1}^{\le 1}$ &
                $\varepsilon_{> 0.01}^{\le 0.1}$ &
                $\varepsilon_{> 0.001}^{\le 0.01}$
\\
\hline
64   & 23950   &  4792  &  16961   &  1907  &    270 \\
128  & 23950   &     0  &    122   &  8781  &  14595 \\
256  & 23950   &     0  &      0   &    64  &   7279 \\
512  & 23950   &     0  &      0   &     0  &    114 \\
\hline
\end{tabular}
\end{table}

\begin{table}[!]
\caption{\label{tab:ss}Convergence of matrix elements of the
 spin-spin interaction of the Argonne $v_{18}$
potential calculated using the Moshinsky transformation-brackets
approach outlined in Sec.~\ref{ops}.
Similar to table~\ref{tab:dr}.}
\centering
\begin{tabular}{rcrrrr}
\hline
NGAUS & NME & $\varepsilon_{> 1}$ &
                $\varepsilon_{> 0.1}^{\le 1}$ &
                $\varepsilon_{> 0.01}^{\le 0.1}$ &
                $\varepsilon_{> 0.001}^{\le 0.01}$
\\
\hline
64   & 43796  &  14168  &  20877  &   8237  &    476 \\
128  & 43796  &    172  &    559  &  16236  &  24244 \\
256  & 43796  &      0  &    172  &    725  &  14238 \\
512  & 43796  &      0  &      0  &    179  &    758 \\
\hline
\end{tabular}
\end{table}


\section{Results and Discussions}
\label{discuss}

To study the numerical accuracy of the approach described in the
previous sections, we will consider all possible \emph{ph}-coupled
matrix elements, NME, of the Argonne $v_{18}$ potential that can be
formed in a $^{16}$O-like model space of single-particle
harmonic-oscillator wave functions, $\mathcal{HO}_{nl}$, with $l \le
6$ and $n \le 6$: the states involving $\mathcal{HO}_{10}$ and
$\mathcal{HO}_{11}$, i.e. $(nlj)=\{
(10\frac{1}{2}),(11\frac{1}{2}),(11\frac{3}{2}) \}$, are taken to be
\emph{hole} (occupied) states, and all others are \emph{particle}
(unoccupied) states.

We will begin by studying the convergence of the matrix elements
calculated using the Moshinsky transformation-brackets approach
outlined in Sec.~\ref{ops}.
Up to intrinsic round-off errors, the numerical accuracy of the
matrix element calculation is tied to the calculation of the radial
part of the matrix element, given in Eq.~(\ref{radmat}). While this
is not the only possible way to calculate this integral, we find it
instructive to perform this integral using the Gauss-Hermite
quadrature formula. This approach is applicable because the
asymptotic form of the product of two harmonic-oscillator wave
functions is Gaussian, and the potential itself falls to zero, as
$r$ goes to infinity. To study the convergence of the matrix
elements with NGAUS, the number of abscissas and weights in the
Gauss-Hermite quadrature set, it is convenient to introduce the
parameter $\varepsilon_{>x}^{\le y}$, which represents the number of
matrix elements characterized by a
percentage change in the numerical value of the matrix element
between $x\%$ and $y\%$ when doubling the number of grid points, NGAUS.


\begin{table}[t]
\caption{\label{tab:te}Convergence of matrix elements of the
 tensor interaction of the Argonne $v_{18}$
potential calculated using the Moshinsky transformation-brackets
approach outlined in Sec.~\ref{ops}.
Similar to table~\ref{tab:dr}.}
\centering
\begin{tabular}{rcrrrr}
\hline
NGAUS & NME & $\varepsilon_{> 1}$ &
                $\varepsilon_{> 0.1}^{\le 1}$ &
                $\varepsilon_{> 0.01}^{\le 0.1}$ &
                $\varepsilon_{> 0.001}^{\le 0.01}$
\\
\hline
64  &  45853  &  16337  &  20967  &   7729  &    751 \\
128 &  45853  &   1907  &  12680  &   6663  &   3845 \\
256 &  45853  &    121  &   1851  &  12795  &   6667 \\
512 &  45853  &      6  &    147  &   2111  &  13314 \\
\hline
\end{tabular}
\end{table}

\begin{table}[!]
\caption{\label{tab:ls}Convergence of matrix elements of the
 spin-orbit interaction of the Argonne $v_{18}$
potential calculated using the Moshinsky transformation-brackets
approach outlined in Sec.~\ref{ops}.
Similar to table~\ref{tab:dr}.}
\centering
\begin{tabular}{rcrrrr}
\hline
NGAUS & NME & $\varepsilon_{> 1}$ &
                $\varepsilon_{> 0.1}^{\le 1}$ &
                $\varepsilon_{> 0.01}^{\le 0.1}$ &
                $\varepsilon_{> 0.001}^{\le 0.01}$
\\
\hline
64  &  41255  &   7931  &  20655  &  10778  &   1724 \\
128 &  41255  &     10  &    108  &    556  &   5059 \\
256 &  41255  &      0  &      8  &     78  &    428 \\
512 &  41255  &      0  &      0  &      6  &     66 \\
\hline
\end{tabular}
\end{table}


Results for the 7~types of operators discussed in the Argonne
$v_{18}$ potential are presented in Tables~\ref{tab:dr}--\ref{tab:ls2}. We find that the
calculation of the matrix-elements converges relatively fast with
the number of points in the Gauss-Hermite quadrature set, NGAUS,
with the one notable exception of the matrix element of the tensor
interaction. In order to converge the tensor matrix element we
require a large set of Gauss-Hermite abscissas. Obtaining such a set
is in fact a nontrivial endeavor, and we have described in~\ref{GH}
a practical approach to achieve this. The difficulty in
converging numerically the tensor matrix element is related to the
fact that the radial shape of the tensor interaction has a longer
range than the other components of the Argonne $v_{18}$ potential.
Hence the deviations from the $e^{-x^2}$ tail of the integrand in
Eq.~(\ref{radmat}) are more pronounced in the case of the tensor
interaction. A larger cutoff and a finer Gauss-Hermite grid is
needed for an accurate numerical evaluation, which in turn leads to
a larger NGAUS number of points.


\begin{table}[t]
\caption{\label{tab:l2}Convergence of matrix elements of the
 $\ell^2$ interaction of the Argonne $v_{18}$
potential calculated using the Moshinsky transformation-brackets
approach outlined in Sec.~\ref{ops}.
Similar to table~\ref{tab:dr}.}
\centering
\begin{tabular}{rcrrrr}
\hline
NGAUS & NME & $\varepsilon_{> 1}$ &
                $\varepsilon_{> 0.1}^{\le 1}$ &
                $\varepsilon_{> 0.01}^{\le 0.1}$ &
                $\varepsilon_{> 0.001}^{\le 0.01}$
\\
\hline
64  &  30650  &   5560  &  15907  &   7636  &   1421 \\
128 &  30650  &      0  &    130  &    254  &   7045 \\
256 &  30650  &      0  &      0  &     82  &    158 \\
512 &  30650  &      0  &      0  &      0  &     82 \\
\hline
\end{tabular}
\end{table}

\begin{table}[!]
\caption{\label{tab:l2s}Convergence of matrix elements of the
 $\ell^2 (\sigma_1 \cdot \sigma_2)$ interaction
of the Argonne $v_{18}$ potential
calculated using the Moshinsky transformation-brackets
approach outlined in Sec.~\ref{ops}.
Similar to table~\ref{tab:dr}.}
\centering
\begin{tabular}{rcrrrr}
\hline
NGAUS & NME & $\varepsilon_{> 1}$ &
                $\varepsilon_{> 0.1}^{\le 1}$ &
                $\varepsilon_{> 0.01}^{\le 0.1}$ &
                $\varepsilon_{> 0.001}^{\le 0.01}$
\\
\hline
64  &  44261  &   8280  &  22408  &  11709  &   1715 \\
128 &  44261  &      2  &     89  &    765  &   8686 \\
256 &  44261  &      0  &      0  &      2  &    163 \\
512 &  44261  &      0  &      0  &      0  &      2 \\
\hline
\end{tabular}
\end{table}

\begin{table}[!]
\caption{\label{tab:ls2}Convergence of matrix elements of the
 quadrupole spin-orbit interaction in the Argonne $v_{18}$
potential calculated using Moshinsky transformation-brackets
approach outlined in Sec.~\ref{ops}.
Similar to table~\ref{tab:dr}.
}
\centering
\begin{tabular}{rcrrrr}
\hline
NGAUS & NME & $\varepsilon_{> 1}$ &
                $\varepsilon_{> 0.1}^{\le 1}$ &
                $\varepsilon_{> 0.01}^{\le 0.1}$ &
                $\varepsilon_{> 0.001}^{\le 0.01}$
\\
\hline
64  &  46726  &  17572  &  22013  &   6245  &    787 \\
128 &  46726  &     31  &    408  &   5549  &  21721 \\
256 &  46726  &      0  &     10  &    106  &   1412 \\
512 &  46726  &      0  &      0  &      6  &     49 \\
\hline
\end{tabular}
\end{table}


We conclude that a parameter value, NGAUS=512, is large enough to
assure a good accuracy of all components of the Argonne $v_{18}$
matrix elements, including the tensor part of the interaction, as
indicated by the sharp decrease in the parameter
$\varepsilon_{>0.1}$ for NGAUS=512. This statement is supported by
results of an independent calculation, to be discussed next.


\begin{table}[!]
\caption{\label{tab:jj}Convergence of matrix elements of the various
components of the Argonne $v_{18}$ potential calculated using
the approach outlined in Sec.~\ref{app:jjme_app}. Calculations were
performed for values of the number of abscissas and weights in the
Gauss-Hermite quadrature set, NGAUS, equal to 32, 64, 96 and 128.
Here we study the evolution of the parameter $\varepsilon_{>x}^{\le
y}$, which indicates how many matrix elements out of the total number of matrix
elements, NME, that can be formed in a $^{16}$O-like model space of
harmonic-oscillator single-particle wave functions,
$\mathcal{R}_{nl}$, with $l \le 6$ and $n \le 6$, are characterized
by a relative error between $x\%$ and $y\%$ when the number of grid
points,NGAUS, changes.
Calculations for larger values of NGAUS, (NGAUS=96 or 128 where
appropriate) have resulted in zero values of the parameter
$\varepsilon_{>x}^{\le y}$.}
\centering
\begin{tabular}{lccrrrr}
\hline
Operator & NGAUS & NME & $\varepsilon_{> 1}$ &
                $\varepsilon_{> 0.1}^{\le 1}$ &
                $\varepsilon_{> 0.01}^{\le 0.1}$ &
                $\varepsilon_{> 0.001}^{\le 0.01}$
\\
\hline
                                $1$ & 64 & 23950  &      0   &    20  &    411 &   17243 \\
            $\sigma_i\cdot\sigma_j$ & 64 & 43796  &      0   &   137  &   1425 &   22048 \\
                           $S_{ij}$ & 64 & 45853  &  17844   &  4704  &   3795 &   17892 \\
                           $S_{ij}$ & 96 & 45853  &      0   &     4  &     23 &     378 \\
                     $\ell \cdot S$ & 64 & 41255  &     10   &   109  &    636 &    5191 \\
                           $\ell^2$ & 64 & 30650  &      0   &   115  &    257 &    3011 \\
$\ell^2 \, (\sigma_i\cdot\sigma_j)$ & 64 & 44261  &      0   &     9  &    294 &    3659 \\
                 $(\ell \cdot S)^2$ & 64 & 46726  &     22   &   230  &   2971 &   16467 \\
\hline
\end{tabular}
\end{table}

\begin{table}[t]
\caption{\label{tab:jj-mosh}Comparison of the matrix elements of the
7~operators in the Argonne $v_{18}$ potential calculated using
the center-of-mass separation of variables for harmonic-oscillator
wave functions such as described in Sec.~\ref{ops}
and the general procedure outlined in Sec~\ref{app:jjme_app}. Here, the
parameter $\varepsilon_{>x}^{\le y}$ reflects the changes between
the Moshinsky-transformation based result (NGAUS=512) and the matrix
elements calculated using the approach presented in Sec.~\ref{app:jjme_app} for
NGAUS=96.}
\centering
\begin{tabular}{lcrrrr}
\hline
Operator & NME & $\varepsilon_{> 1}$ &
                $\varepsilon_{> 0.1}^{\le 1}$ &
                $\varepsilon_{> 0.01}^{\le 0.1}$ &
                $\varepsilon_{> 0.001}^{\le 0.01}$
\\
\hline
                                $1$ & 23950  &      0  &      0  &      0  &      0 \\
            $\sigma_i\cdot\sigma_j$ & 43796  &      0  &      0  &      0  &    188 \\
                           $S_{ij}$ & 45853  &      0  &      8  &    187  &   3365 \\
                     $\ell \cdot S$ & 41255  &      0  &      0  &      0  &      2 \\
                           $\ell^2$ & 30650  &      0  &      0  &      0  &      0 \\
$\ell^2 \, (\sigma_i\cdot\sigma_j)$ & 44261  &      0  &      0  &      0  &      0 \\
                 $(\ell \cdot S)^2$ & 46726  &      0  &      0  &      0  &      4 \\
\hline
\end{tabular}
\end{table}


Next, we will compare the calculation of the Argonne $v_{18}$ matrix elements
using the Moshinsky transformation-brackets approach described in Sec.~\ref{ops} with the more general calculation of the same matrix elements
in the framework outlined in Sec.~\ref{app:jjme_app}.
With the exception of unavoidable round-off errors, the source of possible
numerical loss of accuracy in the calculation of the matrix elements
presented here is linked to the calculation of the integrals
discussed in Sec.~\ref{seq:radial}. First, we are concerned with the
integrals in Eqs.~(\ref{3_6}) and~(\ref{3_13}): Given that the
integrand vanishes for large values of the integration variables,
and in order to maintain the approach general, we will evaluate
these integrals by choosing a large cutoff and discretizing the
integrand on a fine grid. Because these integrals only have to be
performed once in the beginning of the calculation, and that the
results can be stored on disk for future reference, we have chosen a
radial cutoff of 25~fm and a momentum cutoff of 30~fm$^{-1}$ in
Eqs.~(\ref{3_6}) and~(\ref{3_13}), respectively, together with
uniform grids of 8000~points, and have performed these integrals
using the Simpson rule.

Second, we have to perform the integrals involving the radial
parts of the single-particle wave functions, as seen in
Eqs.~(\ref{3_10}) and~(\ref{3_12}). For an arbitrary functional form
of the single-particle wave functions, these calculations can become
computationally expensive, as they may have to be repeated for the
calculation of each and every matrix element. For the case of a
linear combination of harmonic-oscillator single-particle basis, these
integrals can be performed using Gauss-Hermite quadrature formulas.

We study the convergence of the matrix elements
with the number of abscissas and
weights in the Gauss-Hermite quadrature set, NGAUS, by computing the
value of the parameter $\varepsilon_{>x}^{\le y}$, which represents
the number of matrix elements characterized by a
percentage change in the numerical value of the matrix element
between $x\%$ and $y\%$ as the number of grid points, NGAUS, changes.
We have performed calculations using 4~sets of Gauss-Hermite
quadrature points, corresponding to values of the NGAUS parameter of
32, 64, 96, and 128.
Results for the 7~types of operators part of the Argonne $v_{18}$ potential are
presented in Table~\ref{tab:jj}. We find that the calculation of the
matrix-elements is fully converged for a value of the NGAUS
parameter, NGAUS=96, for all components of the Argonne $v_{18}$
potential, with the exception of the tensor interaction, which
requires a larger set of NGAUS=128 Gauss-Hermite quadrature points.
The slower convergence of the tensor-interaction matrix element
follows the trend observed in the case of the calculation based on
the Moshinsky transformation brackets, and is due to the fact
that the tensor interaction in the Argonne $v_{18}$ potential has a
longer range and more structure than the other components of the
interaction.

In order to ascertain the numerical accuracy and correctness of the
approach presented in Sec.~\ref{app:jjme_app}, we have also compared  with
results of the calculation of the matrix elements using the
Moshinsky transformation brackets discussed earlier. Results are
displayed in Table~\ref{tab:jj-mosh}. Here, the parameter
$\varepsilon_{>x}^{\le y}$ reflects the changes between the
Moshinsky transformation-brackets result (NGAUS=512) and the matrix
elements calculated using the approach discussed in Sec.~\ref{app:jjme_app} for
NGAUS=96. The two calculations lead to numerically identical matrix
elements, which is quite remarkable since the two approaches are
very different. This validates both theoretical approaches.

In summary, in this paper we discussed two approaches for the calculation of
two-body matrix elements of the Argonne $v_{18}$ potential.
The first approach is only applicable to the specific case of a
harmonic-oscillator single-particle wave function representation.
In this case, the matrix elements are calculated
using the Talmi transformation (implemented numerically
using the Moshinsky transformation brackets) which allows for the separation of the center-of-mass and relative coordinates degrees of freedom.
Integrals involving the radial part of
the potential were performed using Gauss-Hermite quadrature
formulas, and convergence was achieved for sets of at least
512~Gauss points. This procedure was designed to validate the
calculation of matrix elements of the Argonne $v_{18}$
potential using an approach suitable for an arbitrary functional form of the
single-particle wave functions.
The latter approach represents the main thrust of this paper.
This general framework is suitable for the calculation of matrix elements
involving a representation of the single-particle wave functions
given by linear combinations of harmonic-oscillator
wave functions such as in Refs.~\cite{exps-1,exps-2,exps-3},
and/or the two-scale functional representation for the hole and particle
sides of the spectrum such as outlined in Ref.~\cite{cce-a1}.
For a model space represented in terms of harmonic-oscillator
wave functions, results obtained using the approaches discussed in Sec.~\ref{ops} and Sec.~\ref{app:jjme_app} are shown to be identical within numerical accuracy.


\appendix


\section{Generator of large-$N$ Gauss-Hermite quadrature sets}
\label{GH}

The algorithm used to generate the Gauss-Hermite quadrature
abscissas and weights represents an adaption of the algorithm
described in Ref.~\cite{nr}.

The Gauss-Hermite quadrature formula allows for the efficient
numerical computation of the integral
\begin{equation}
   \int_{-\infty}^\infty \ \mathrm{d}x \
   f(x) \ e^{-x^2} \ \approx \
   \sum_{j=1}^N \ w_j \ f(x_j)
   \>,
\label{gauss}
\end{equation}
where $x_j$ are the roots of the $H_N(x)$ Hermite polynomial. The
$N$-points Gauss-Hermite becomes exact provided that $f(x)$ is a
polynomial. In order to avoid the overflow problems related to the
standard recurrence relation of the Hermite polynomials, i.e.
\begin{equation}
   H_{j+1}(x) \ = \
   2\,x \ H_j(x) \ - \ 2\,j \ H_{j-1}(x)
   \>,
\end{equation}
one uses instead the the orthonormal set of polynomials, $\tilde
H_j(x)$, which are generated using the recurrence relation
\begin{eqnarray}
   &&
   \tilde H_{-1} = 0 \>,
   \quad
   \tilde H_{0} = \pi^{-1/4} \>,
   \\ \nonumber &&
   \tilde H_{j+1}(x)
   \ = \
   x \ \sqrt{\frac{2}{j+1}} \ \tilde H_j(x) \ - \ \sqrt{\frac{j}{j+1}} \ \tilde H_{j-1}(x)
   \>.
\end{eqnarray}
Then, the formula for the Gauss-Hermite weights is
\begin{equation}
   w_j \ = \ \frac{2}{|\tilde H_j'(x_j)|^2}
   \>,
\end{equation}
where $\tilde H_j'(x)$ denotes the derivative of the polynomial
$\tilde H_j(x)$, which is calculated as
\begin{equation}
   \tilde H_j'(x) \ = \ \sqrt{2j} \ \tilde H_{j-1}(x)
   \>.
\end{equation}
The roots of the Hermite polynomial $\tilde H_N(x)$ are symmetric
about the origin, so we only have to calculate half of them, e.g.
the positive ones. The roots are estimated using the formulas put
forward by Szeg\"o for the largest root of the Hermite polynomial
$H_N(x)$~\cite{szego}, i.e.
\begin{equation}
   x_N^{[0]} \ \simeq \
   \sqrt{2N+1} \ - \ 1.85575 \, (2N+1)^{-1/6}
   \>,
\end{equation}
and by Stroud and Secrest~\cite{stroud} for the second-largest root,
\begin{equation}
   x_{N-1}^{[0]} \ \simeq \
   x_N \ - \ 1.14 \, N^{-.426} / x_N
   \>,
\end{equation}
the third-largest root
\begin{equation}
   x_{N-2}^{[0]} \ \simeq \
   1.86 \, x_{N-1} \ - \ 0.86 \, x_N
   \>,
\end{equation}
the fourth-largest root
\begin{equation}
   x_{N-3}^{[0]} \ \simeq \
   1.91 \, x_{N-2} \ - \ 0.91 \, x_{N-1}
   \>,
\end{equation}
and all other positive roots
\begin{equation}
   x_j^{[0]} \ \simeq \
   2 \, x_{j-1} \ - \ x_{j-2}
   \>.
\end{equation}
Subsequently, a root-finding algorithm is employed to improve these
solutions. For added accuracy, all calculations are performed in
quadruple precision, even though the Gauss-Hermite quadrature sets
are ultimately used in double precision.

Press \emph{et al.}~\cite{nr} propose the commonly-used
\emph{Newton's method} to achieve this task. In this approach, the
$i^{\mathrm{th}}$ approximation of the root $x_j$ is obtained as
\begin{equation}
   x_j^{[i]} \ = \ x_j^{[i-1]} \ - \ \frac{H_N(x_j^{[i-1]})}{H_N'(x_j^{[i-1]})}
   \>.
\end{equation}
In our experience, this procedure is able to successfully produce
all $N$-points Gauss-Hermite quadrature sets of abscissas and
weights for $N\lesssim 200$. For larger sets, one must replace the
root finder and use \emph{Muller's method} instead (see page 364 in
Ref.~\cite{nr}.) This method uses a quadratic interpolation among
the 3 most recent estimates of the solution. To begin the search, we
have chosen the first 3 guesses of the solution as $x_{j}^{[0]}$,
given by the Szeg\"o's or Stroud's approximations described above,
together with the coordinates $x_{j}^{[0]} \pm 0.1$. It is important
for this root-finding algorithm to be allowed to carry out the
search in the \emph{complex} plane, even though the roots of the
Hermite polynomial are real. Muller's method allows for the
calculation of Gauss-Hermite quadrature sets with $N\lesssim 600$.

To further improve the algorithm and reach higher values of~$N$, it
has been noticed that the coordinate dropped after the \emph{first}
estimate performed using Muller's method, must be the one which is
the farthest away from this estimate. This allows for the
calculation of Gauss-Hermite quadrature sets with $N\lesssim 1050$,
at which point we are very close to the natural numerical barrier
encountered in numerical computations performed using 32-bit CPUs,
even when using quadruple precision. Because, these sets where large
enough to provide convergence for the purpose of our matrix-elements
calculations, and because our calculations are carried out entirely
on 32-bit platforms, we have not investigated this matter further.

Because of the symmetry of the Gauss-Hermite quadrature set and the
fact that our integrals involve only the radial coordinate in a
spherical coordinates, we define the parameter NGAUS as half the
value of the integer $N$ in Eq.~(\ref{gauss}).


\section{Reduced matrix elements}

In this appendix we list commonly used reduced matrix elements,
relevant to the calculations described in this paper.


\subsection{$C^{(k)}(\hat r)$}

The reduced matrix element of the \emph{unnormalized} spherical
harmonic is (\cite{Edmonds}, Eq.~5.4.6)
\begin{equation}
   \langle l_1 \; \| \; \ck{k}{r} \; \| \; l_2 \rangle \>
   \ = \
\label{ck_red}
   \hat l_2 \ \cg{l_2}{0}{k}{0}{l_1}{0}
   \>,
\end{equation}
with the selection rules $\ell_1+\ell_2+k = even$, and $|l_1-l_2|
\le k \le l_1+l_2$.


\subsection{$\bigl [ C^{(k_1)}(\hat r) \otimes \nabla \bigr ]^{(k_2)}$}

Using (\cite{Edmonds}, 7.1.1) we obtain
\begin{align}
   &
   \langle n_1 l_1 \|
   \bigl [ C^{(k_1)}(\hat r) \otimes \nabla \bigr ]^{(k_2)}
   \| n_2 l_2 \rangle
   \\ \nonumber &
   =
   (-)^{l_1+l_2+k_2} \, \hat k_2
   \sum_l \wsj{k_1}{1}{k_2}{l_2}{l_1}{l}
   \langle l_1 \| C^{(k_1)}(\hat r) \| l \rangle \,
   \langle n_1 l \| \nabla \| n_2 l_2 \rangle
   \>,
\end{align}
subject to the selection rules: $|k_1-1| \le k_2 \le k_1+1$. The
reduced matrix elements of the gradient operator are obtained from
the gradient formula (see page 79 of Ref.~\cite{Edmonds}). The only
nonzero components are
\begin{align}
   &
   \langle n_1 (l_1=l_2-1) \| \nabla
                                 \| n_2 l_2 \rangle
   =
   - \sqrt{l_2}
   \int_0^\infty [ r^2 \, {\rm d}r ]
   R_{n_1 l_1}(r)
   \Bigl [
       \Bigl (
       \frac{d}{dr} + \frac{l_2+1}{r}
       \Bigr )
       R_{n_2l_2}(r)
   \Bigr ]
   \>,
\end{align}
and
\begin{align}
   &
   \langle n_1 (l_1=l_2+1) \| \nabla \| n_2 l_2 \rangle
   =
   \sqrt{l_2+1}
   \int_0^\infty \, [ r^2 \ {\rm d}r ] \,
   R_{n_1 l_1}(r) \,
   \Bigl [
       \Bigl (
       \frac{d}{dr} - \frac{l_2}{r}
       \Bigr ) \
       R_{n_2l_2}(r)
   \Bigr ]
   \>.
\end{align}


\subsection{$\bigl [ C^{(k_1)}(\hat r) \otimes [ \nabla \otimes \nabla ]^{(j)} \bigr ]^{(k_2)}$}

Using (\cite{Edmonds}, 7.1.1) we obtain
\begin{align}
   &
   \langle n_1 l_1 \|
   \bigl [ C^{(k_1)}(\hat r) \otimes [ \nabla \otimes \nabla ]^{(j)} \bigr ]^{(k_2)}
   \| n_2 l_2 \rangle
   \\ \nonumber & \
   =
   (-)^{l_1+l_2+k_2} \, \hat k_2 \
   \sum_l \wsj{k_1}{j}{k_2}{l_2}{l_1}{l}
   \langle l_1 \| C^{(k_1)}(\hat r) \| l \rangle \,
   \langle n_1 l \| [ \nabla \otimes \nabla ]^{(j)} \| n_2 l_2 \rangle
   \>,
\end{align}
with the selection rules: $|\ell-1| \le \kappa \le \ell+1$. The
allowed reduced matrix elements are
\begin{align}
   &
   \langle n_1 (l_1 = l_2-2) \|
       [ \nabla \otimes \nabla ]^{(j)}
   \| n_2 l_2 \rangle
   =
   \sqrt{l_2(l_2-1)}
   \wsj{1}{1}{j}{l_2}{l_2-2}{l_2-1} \
   \\ \nonumber & \ \times
   (-)^j \hat j
   \int_0^\infty [ r^2 \ {\rm d}r ]
   R_{n_1l_1}(r)
   \Bigl \{
       \Bigl [
       \frac{d^2}{dr^2} + \frac{2l_2+1}{r} \frac{d}{dr} + \frac{(l_2-1)(l_2+1)}{r^2}
       \Bigr ]
       R_{n_2l_2}(r)
   \Bigr \}
   \>,
\end{align}
and
\begin{align}
   &
   \langle n_1 (l_1 = l_2) \|
              [ \nabla \otimes \nabla ]^{(j)}
   \| n_2 l_2 \rangle
   \nonumber \\ & =
   -
   \Biggl [
   l_2 \wsj{1}{1}{j}{l_2}{l_2}{l_2-1} +
   (l_2+1) \wsj{1}{1}{j}{l_2}{l_2}{l_2+1}
   \Biggr ]
   \\ \nonumber & \ \times
   (-)^j \hat j
   \int_0^\infty \ [ r^2 \ {\rm d}r ] \
   R_{n_1l_1}(r) \ \phi(r) \
   \Bigl \{
       \Bigl [
       \frac{d^2}{dr^2} + \frac{2}{r} \frac{d}{dr} - \frac{l_2(l_2+1)}{r^2}
       \Bigr ] \
       R_{n_2l_2}(r)
   \Bigr \}
   \>,
\end{align}
and
\begin{align}
   &
   \langle n_1 (l_1 = l_2+2) \|
       [ \nabla \otimes \nabla ]^{(j)}
   \| n_2 l_2 \rangle
   = \
   \sqrt{(l_2+1)(l_2+2)} \
   \wsj{1}{1}{j}{l_2}{l_2+2}{l_2+1} \
   \nonumber \\ & \ \times
   (-)^j \hat j
   \int_0^\infty \, [ r^2 \ {\rm d}r ] \
   R_{n_1l_1}(r) \
   \Bigl \{
       \Bigr [
       \frac{d^2}{dr^2} - \frac{2l_2+1}{r} \frac{d}{dr} + \frac{l_2(l_2+2)}{r^2}
       \Bigr ] \
       R_{n_2l_2}(r)
   \Bigr \}
   \>.
\end{align}


\ack

This work was performed in part under the auspices of the U.S. Department of Energy.
The author acknowledges useful conversations with J.H. Heisenberg and J.F. Dawson.


\end{document}